\documentclass[preprint,12pt]{elsarticle}

\usepackage{subfigure}
\usepackage{booktabs}
\usepackage{multirow}
\usepackage{booktabs,array}
\usepackage{graphicx}
\usepackage{CJKutf8}
\usepackage{xcolor}
\usepackage[normalem]{ulem}
\usepackage{hyperref}
\usepackage{blindtext, rotating}
\newdimen\imageheight

\journal{Computers in Human Behavior}

\begin{document}

\begin{frontmatter}



\title{Exploring Emotional and Social Dynamics in Mobile Usage During Home Confinement}

\author[aff_c,aff_a,aff3]{Nan Gao\fnref{cofirst}}
\ead{nan.gao@nankai.edu.cn}
\affiliation[aff_c]{organization={Nankai  University}, 
            city={Tianjin},
            postcode={300350}, 
            country={China}}
\affiliation[aff_a]{organization={University of New South Wales (UNSW)}, 
            city={Sydney},
            postcode={1466}, 
            country={Australia}}

\author[aff3]{Sam Nolan\fnref{cofirst}}
\ead{samnolan555@gmail.com}
\affiliation[aff3]{organization={RMIT University},
            city={Melbourne},
            postcode={3000}, 
            country={Australia}}

\author{Kaixin Ji\fnref{cofirst}}
\ead{kaixin.ji@student.rmit.edu.au}

\author[aff3]{Shakila Khan Rumi}
\ead{shakilakhanrumi@gmail.com}

\author[aff_g]{Judith Simone Heinisch}
\ead{judith.heinisch@comtec.eecs.uni-kassel.de}
\affiliation[aff_g]{organization={University of Kassel},
            city={Kassel},
            postcode={34131}, 
            country={Germany}}

\author[aff_g]{Christoph Anderson}
\ead{c\_anderson@posteo.de}

\author[aff_g]{Klaus David}
\ead{david@uni-kassel.de}

\author[aff_a]{Flora D. Salim\corref{cor1}}
\ead{flora.salim@unsw.edu.au}

\fntext[cofirst]{Equally contributing first authors.}
\cortext[cor1]{Corresponding author}

\begin{abstract}
Home confinement, a situation experienced by individuals for reasons ranging from medical quarantines, rehabilitation needs, disability accommodations, and remote working, is a common yet impactful aspect of modern life. While essential in various scenarios, it can profoundly influence mental well-being and digital device usage. Using the COVID-19 lockdown as a case study, this research explores the emotional and social effects of home confinement on mobile device usage. We conducted an in-situ study with 32 participants, analyzing three weeks of mobile usage data to assess emotional well-being and social dynamics in restricted environments. We found that app usage patterns serve as strong indicators of emotional states, offering insights into how digital interactions can reflect and influence well-being during isolation. This study highlights the potential for developing interventions and support systems for individuals in long-term home confinement, including those with chronic illness, recovery needs, or permanent remote work situations.
\end{abstract}




\end{frontmatter}

\newcommand{\kx}[1]{\textcolor{black}{#1}}

\newcommand{\gn}[1]{\textcolor{black}{#1}}


\section{Introduction}

Home confinement, defined as the restriction of movements within the home \cite{narici2021impact} due to illness, recovery, disability, or other personal circumstances, is a common phenomenon that everyone may experience during their lives \cite{kitwood1992towards,cay1972psychological}. 
Research has shown that such confinement can adversely affect mental well-being, leading to stress, anxiety, and depression \cite{bartoszek2020mental,rodriguez2021psychological}. For instance, Bartoszek et al. \cite{bartoszek2020mental} found that quarantined individuals experienced heightened levels feelings of depression, insomnia, loneliness, and fatigue. Understanding these impacts is essential for developing effective targeted interventions to support individuals during such periods.

Traditionally, mental well-being during home confinement has been assessed using survey-based instruments \cite{rodriguez2021psychological,gaeta2021impact}, which can be time-consuming, labour-intensive, and prone to biases such as social desirability and reflection bias \cite{gao2021investigating,paulhus1991measurement}. Recent advances in mobile sensing, however, offer a more objective, unobtrusive and real-time alternative \cite{rashid2020socialanxiety, chow2017using}.  While methods relying on physical movement data (e.g., GPS and accelerometers) have been widely used for mental wellbeing modelling such as social anxiety, affect, depression and social isolation \cite{rashid2020socialanxiety,chow2017using}, they are less effective in the context of indoor confinement, where movement patterns (e.g., step counts, visited locations) are limited \cite{rashid2020socialanxiety}. This study shifts focus to more granular data, such as app usage patterns, to provide deeper insights into emotional well-being during confinement.

The diversity of individual circumstances complicates the study of home confinement. Factors such as the reason for confinement, personal health conditions, and social environment can vary greatly, complicating the creation of a standardized model. However, the COVID-19 lockdown provided a unique, global and natural context in which the majority of individuals experienced similar restrictions, with a marked increased reliance on digital technologies \cite{pandey2020impact,sebire2020coronavirus}. This shared experience allows for a more uniform analysis of how mobile device usage reflects emotional states, social roles, and work-life dynamics during such periods. Especially, while this study uses the lockdown as a descriptive case study \cite{yin2009case}, the insights gained may also be applicable to other forms of home confinement, such as those experienced by individuals with chronic illnesses, in post-recovery rehabilitation, or those adapting to permanent remote work arrangements.

In this study, we explore the relationship between mobile device interactions, emotional well-being, and social roles during home confinement, using data collected during the COVID-19 lockdown as a case study. We focus on two key aspects of well-being: emotions and social roles, as both are critical to mental health \cite{de2013beliefs, nordenmark2004multiple}. By utilizing the \textit{Valence} and \textit{Arousal} dimensions from the \textit{Circumplex Model of Affect} \cite{russell1980circumplex} and \textit{Social Role} from the \textit{Role Theory} \cite{biddle1986recent}, we aim to provide a comprehensive understanding of how mobile device usage during lockdown reflects emotional and social dynamics. Our research seeks to address the following questions: 
\textbf{RQ1.} \textit{How are emotions and social roles during home confinement perceived through self-reports and inferred from app usage patterns?}
\textbf{RQ2.} \textit{Can mobile usage data be used to model emotional well-being during home confinement, and which features are most predictive?}

To answer these questions, we conducted a three-week data collection during the COVID-19 lockdown in Melbourne, Australia, involving 32 participants.
We analyzed perceived emotional states, social roles, and work-life balance through \textit{Experience Sampling Method} (ESM) \cite{beal_esm_2015} and end-of-day surveys. We also proposed a novel method, \textit{App-based Information Gain Temporal Segmentation} (AIGTS), to extract user activities from app usage records by identifying segments and transition times between apps during the day. The resulting insights offer a deeper understanding of how app usage during confinement reflects emotional well-being and social dynamics, with implications for developing digital interventions to support individuals in long-term home confinement. 

All scripts for this study are open source.\footnote{Link will be shared after the double-blind review.}
In summary, our contributions are as follows:
\begin{itemize}
    \item We collected a rich dataset from 32 participants during a 3-week COVID-19 home confinement period in Melbourne, Australia. This dataset includes 502,485 records of users' smartphone and desktop usage data, 1,749 ESM responses, and 265 end-of-day surveys. It provides a unique, real-world snapshot of emotional well-being and social dynamics during lockdown.

    \item We analyzed perceived valence, arousal, and social roles, revealing that the lockdown significantly impacted well-being, reducing emotional valence and increasing work hours due to interruptions or blurred work-life boundaries. These findings shed light on the complex emotional landscape during prolonged confinement.
    
    \item We introduced the App-based Information Gain Temporal Segmentation (AIGTS) method, an unsupervised technique for extracting user activities (e.g., video-watching, socializing, deep work) from app usage data. This method autonomously identified 16 distinct user activities, providing granular insights into how digital behaviors relate to emotional well-being.
    
    \item 
    We demonstrated that mobile usage patterns can predict emotional well-being during home confinement. Our extensive experiments revealed key features that are most predictive of emotional states, offering new avenues for developing digital interventions to support individuals during isolation.
    
\end{itemize}

\section{Related work}\label{sec:related}


\subsection{Impact of the Home Confinement on Mental Well-Being}

Home confinement, whether due to medical conditions, disability, or societal events like pandemics, can significantly affect mental well-being. Prolonged isolation, particularly when coupled with limited physical activity and disrupted social interactions, is linked to increased risks of anxiety, depression, stress, and loneliness \cite{talevi2020mental, choi2020taylor, brooks2020psychological}. The psychological impact is often compounded by the loss of routine and control, leading to emotional disturbances such as irritability, confusion, and anger \cite{reynolds2008understanding, lee2005experience}. These effects may persist even after restrictions are lifted \cite{jeong2016mental}.


The COVID-19 pandemic underscored these effects on a global scale. The World Health Organization \cite{world_health_organization_mental_2020} reported that the psychological toll of lockdowns and social distancing measures often surpassed the physical health risks. Studies during the pandemic have documented a significant increases in mental health issues, with depression and anxiety becoming widespread \cite{xiong_impact_2020, daly2020longitudinal, killgore2020loneliness}. For example, Khubchandani et al. \cite{khubchandani2021post} found depression (39\%) and anxiety (42\%) surged in the U.S. post-lockdown. Similarly, Fountoulakis et al. \cite{fountoulakis2021self} reported increased anxiety (45\%) and suicidal thoughts (10.4\%) in Greece. Loneliness was also a major issue, especially among those in long-term isolation \cite{killgore2020loneliness}.

Despite extensive research on the mental well-being impacts of various forms of home confinement, much of it is geographically limited or reflects different phases of isolation. This variability makes it challenging to fully understand the broader, long-term psychological effects of home confinement in diverse contexts. To address this gap, \textbf{RQ1} examines how emotions and social roles are experienced during home confinement, using self-reports and app usage patterns to infer these dynamics.

\subsection{Digital Device Usage During Home Confinement}

During periods of home confinement, digital devices play a crucial role in maintaining communication, accessing information, and fulfilling work and social roles. With restricted physical mobility, devices such as smartphones, tablets, and computers become essential for staying connected to social networks and managing day-to-day activities \cite{jo2024influence}. For instance, Ratan et al. \cite{ratan2021smartphone} found that, during home confinement, smartphones became a vital tool for work-related communication and personal connections, often replacing face-to-face interactions.

While many studies have explored digital device usage in general contexts, the COVID-19 lockdown provided a unique opportunity to study digital usage behavior under prolonged confinement. With workplaces and schools shifting to remote environments, the demand for digital communication tools surged \cite{tyagi2021effects, dunham2023impacts, ankenbauer2020engaging}. Video conferencing platforms, social media, and messaging apps saw massive increases in usage \cite{sebire2020coronavirus, pandey2020impact}, and overall digital device use skyrocketed. A global survey found 70\% of internet users reported increased smartphone usage during the lockdown \cite{sebire2020coronavirus}, emphasizing how integral smartphones became in managing both work and personal life.  Similarly, \citet{majumdar2020covid} observed that office workers spent significantly more time on digital devices, including smartphones, desktops, and televisions, during the lockdown. \citet{jonnatan2022mobile} also found that mobile device usage rose during the pandemic, with urban participants reporting higher use and greater concerns about overuse, addiction, and negative health impacts.

In the post-pandemic era, many individuals continue to rely heavily on digital devices due to the persistence of remote work and online communication \cite{lund2021future, jain2022covid}. Studies suggest that behaviors such as prolonged screen time and device overuse, which emerged during the pandemic, have continued into the post-lockdown phase \cite{reopen2021covid}. This highlights the importance of understanding digital device usage patterns in today’s technology-reliant environments, particularly in contexts of home confinement.

\subsection{Inferring Mental Well-Being Using Mobile Sensing Technologies}

Advances in mobile sensing technologies have opened new avenues for understanding human mental well-being, including mood \cite{morshed_prediction_2019, moodexplorer, meegahapola2023generalization}, engagement \cite{gao2020n, huynh2018engagemon,ananthan2024understanding}, and mental health issues like depression and anxiety \cite{chow2017using, rashid2020socialanxiety, wang_tracking_2018}. Many studies have integrated data from multiple sensors to improve prediction accuracy. For example, Gao et al. \cite{gao2020n} used wearable and environmental sensors to predict student engagement, while \citet{wang_tracking_2018} tracked depression using passive mobile sensing and wearables. \citet{jaques_predicting_2015} modeled student mood using physiological signals and mobile behavior.

While multi-source sensing enhances accuracy, there is increasing interest in using mobile sensing data alone, given its convenience and ability to capture a range of behaviors like physical activity, communication patterns, and location \cite{gao2024leveraging}. Studies have shown that mobile usage data can predict psychological states such as anxiety \cite{rashid2020socialanxiety, tlachac2022deprest}, mood \cite{moodexplorer, meegahapola2023generalization}, and emotion \cite{tag2022emotion}. For instance, \citet{moodexplorer} developed an app to detect compound emotions from smartphone usage, and \citet{tlachac2022deprest} used call and text logs to predict anxiety and depression during the pandemic.

Research on app sequences has also provided insights into daily behaviors, such as social roles \cite{anderson_impact_2019} and interruptibility \cite{okoshi_attelia_2014, okoshi_attelia_2015}. \citet{anderson_impact_2019} demonstrated that app sequences correlate with individuals' work and private roles, while \citet{okoshi_attelia_2014,okoshi_attelia_2015} used app sequences to predict activity breakpoints for intelligent notification delivery. However, most studies have focused on identifying breakpoints rather than modeling user activity or predicting contextual variables. Additionally, earlier studies often relied on basic aggregation methods, 
which fail to capture transitions between activities. 
Thus, \textbf{RQ1} also explores emotions and social roles by analyzing app usage patterns.

While mobile sensing technologies have shown potential in predicting well-being, gaps remain in understanding how these tools can be applied specifically during home confinement. Home confinement introduces unique behaviors, such as increased reliance on digital devices and altered perceptions of time, that differ from daily life. Although some studies \cite{tlachac2022deprest, Nepal2022covid,swain2024leveraging} have examined psychological states during the pandemic, they focus on broader environments rather than the specific context of home confinement. Understanding this unique setting could provide valuable insights for other confinement scenarios, such as post-operative recovery, parental leave, or aging.
Therefore, \textbf{RQ2} investigates how mobile sensing can model emotions during home confinement and identify key features for prediction.

\begingroup
\let\clearpage\relax
\section{Study Overview}
\label{sec:data}

We conducted a three-week field study in Melbourne, Australia, from 24 August 2020 to 13 September 2020, during the strictest Stage IV COVID-19 restrictions. This study uses Melbourne's experience as a lens to explore broader home confinement scenarios, such as those due to medical conditions, recovery needs, or remote work. The following section details the background, participants, and data collected.

\subsection{A Case Study of COVID-19 Lockdown in Melbourne, Australia}
\label{subsec: background}

\begin{figure}[htbp!]
    \centering
    \includegraphics[width=0.95\textwidth]{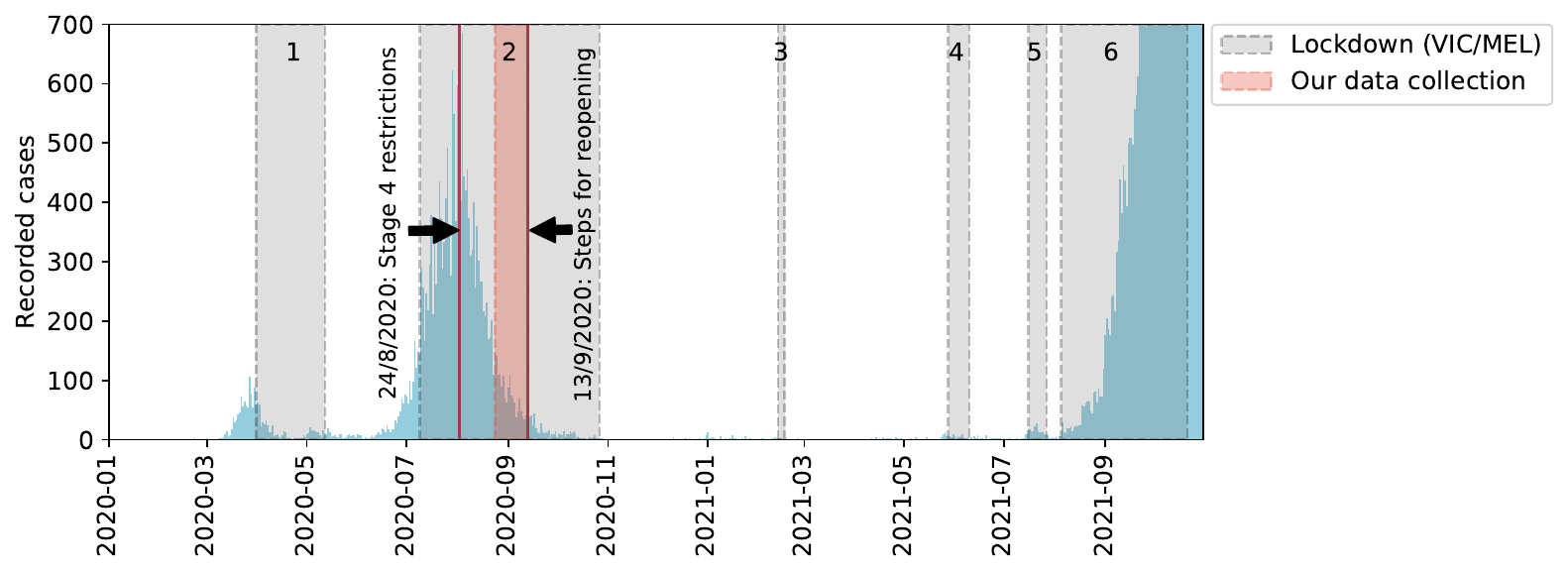}
    \caption{The {periods} of six lockdowns and {our} data collection}
    \label{fig:lockdown}
\end{figure}

To prevent the community spread of COVID-19, Melbourne has been under six lockdowns since March 2020, with a total of 263 days of lockdowns, making it the city with the longest COVID-19 lockdown in the world \cite{matheson2021prematurity, stansfield2022impact}. 
Our data collection was conducted during the second lockdown period with \textit{Stage IV} restrictions (the strictest measures), which lasted from 9 July 2020 to 27 October 2020, with our study conducted from 24 August 2020 to 13 September 2020 (see Figure~\ref{fig:lockdown}). Among the six lockdown periods, this was one of the longest and strictest lockdown periods, during which a state of disaster was declared on 2 August 2020, followed by \textit{Stage IV} restrictions in metropolitan Melbourne \cite{murray-atfield_dunstan_2020}.

Under the \textit{Stage IV} restrictions in force at the time of our data collection, a daily curfew was implemented from 8 PM to 5 AM, forbidding anyone from leaving their homes except for special reasons. Residents were only allowed to shop and exercise within 5 km of their homes, and all students returned to home-based learning. Other restrictions that were previously applied only to specific districts were extended to the entire Melbourne metropolitan area. On 13 September 2020, several restrictions were eased in Melbourne, including reduced curfews and loosened rules on outdoor exercise and social interactions. During home confinement due to the lockdown, people showed a strong reliance on technology, particularly for school and communication \cite{ratan2021smartphone,sebire2020coronavirus}.

\subsection{Procedures}
\label{subsec: par}

Participants were recruited through advertising and comprised of university students, professional workers, academic staff, and health workers. In total, we recruited 32 participants (17 males and 15 females), where most participants were between 18-40 years old (mean = 25.44, median = 23.00, SD = 8.95). Of these, 27 participants provided smartphone data, while 13 contributed desktop data.
Recruitment was primarily through social media platforms \textit{Facebook} and \textit{Discord}, potentially leading to an over-representation of users from these platforms. However, as our study focused on psychological states and digital device usage during COVID-19 home confinement, not social media behaviour, we believe this does not significantly impact the validity of our findings.

Prior to data collection, participants were instructed to install the \textit{Balance for Android} and  \textit{Balance for Desktop} apps\cite{heinisch2022investigating,van_berkel_experience_2017}  on their smartphones and desktops. These apps, designed to record usage behaviors and enable experience sampling, were optimized for minimal computational, energy, and storage consumption.
The apps used  Experience Sampling Method (ESM) to prompt participants with a short questionnaire every 90 minutes during active device use between 7 am and 10 pm. To minimize participant burden, surveys were only triggered during active use. Additionally, the event-based ESM was triggered after 10 minutes of interaction, with follow-ups every 30 minutes. At day's end, participants received an end-of-day survey via email at 7:30 pm.
After the data collection, each participant was compensated with a 25 AUD gift card as a token of appreciation for their time. 

\textbf{Privacy, ethics, and safety.}

Our field study was designed and conducted after carefully reviewing potential risks and benefits, in compliance with the city's COVID-19 safety regulations. The study has been approved by the Human Research Ethics Committee at the authors' university. To ensure privacy, all personal identifying information was anonymized before data analysis. Participants' consent was obtained prior to data collection, with a clear explanation of the study's purpose and the nature of the data being collected. We also implemented strong data security measures, including encryption of sensitive data, to further protect participants’ privacy. Participants were informed of their right to withdraw at any time without penalty, and their data would be excluded from the study upon request.

\begin{table}[ht]
\caption{Overview of the collected data records across devices}
\small
\label{tab:data statistic}
\begin{tabular}{@{}lllll@{}}
\toprule

\textbf{Category}             & \textbf{Device} & \textbf{Total Records} & \textbf{Participants} & \textbf{Data Used for RQs} \\ \midrule
\multirow{3}{*}{\textit{App} \& Both}            & 502,485          & 32                    & RQ1                         \\ \cline{2-5}
                              & Desktop         & 261,911         & 13                    & /                             \\\cline{2-5}
                              & Smartphone      & 240,574         & 27                    & RQ2                         \\\hline
\multirow{3}{*}{\textit{ESM} \& Both}         & 1,749            & 28                    & RQ1                         \\\cline{2-5}
                              & Desktop         & 227             & 10                   &     /                        \\\cline{2-5}
                              & Smartphone      & 1,522            & 25                    & RQ2                         \\\hline
\textit{Ene-of-Day Survey}    & Both            & 265              & 25                    & RQ1                         \\ \bottomrule
\end{tabular}
\end{table}

\begin{table}[]
\centering
\footnotesize
\caption{Detailed overview of data categories, items, and response options}
\begin{tabular}{@{}l@{\hspace{0.15cm}}l@{\hspace{0.15cm}}l@{\hspace{0.15cm}}l@{\hspace{0.15cm}}l@{}}
\toprule
\textbf{Category}                                                                      & \textbf{Items}             & \textbf{Questions / Descriptions}                                                     & \textbf{Options}                   &  \\ \midrule
\multirow{3}{*}{\textit{
\begin{tabular}[c]{@{}l@{}}
\textit{Device}\\ 
\textit{Usage}
\end{tabular}}}                                                          & Application usage           & Foreground application name,  package identifier, category      &     N/A        &  \\
     & Notifications    & Notification arrival time, content (hashed), sender (hashed), relationship    &  N/A           &  \\
                                                                              & Phone's states       &   Screen status, battery, ringer modes, GPS location, physical activities                   &     N/A &  \\ \hline
\multirow{5}{*}{\textit{ESM}}                                                          & Valence           & How did you feel in the last hour? (Unhappy vs Happy)         & 5-Likert scale            &  \\
                                                                              & Arousal           & How did you feel in the last hour? (Calm vs Excited)          & 5-Likert scale            &  \\
                                                                              & Social role       & In the last 15 minutes, I was engaged in:                     & Work/private/both       &  \\
                                                                              & Interruptibility   & I am currently available for other matters:                     & Work/private/both/none &  \\
                                                                              & Task   & The task I managed to progress the most in the last 90 minutes:                     & Multiple choices  \\
                                                                              \hline
\multirow{10}{*}{\begin{tabular}[c]{@{}l@{}}\textit{End-of-Day}\\ \textit{Survey}\end{tabular}} & Productivity      & I feel my day was productive.                                 & 5-Likert scale            &  \\
                                                                              & Interruptibility    & I was continually interrupted today.                          & 5-Likert scale            &  \\
                                                                              & Valence           & Has the lockdown influenced your happiness today?             & 5-Likert scale            &  \\
                                                                              & Arousal           & Has the lockdown influenced your energy today?                & 5-Likert scale            &  \\
                                                                              & Work-life balance & Has the lockdown impacted your work-life balance today?       & 5-Likert scale            &  \\
                                                                              & Valence         & How has the lockdown influenced your happiness today?         & Text (optional)           &  \\
                                                                              & Arousal         & How has the lockdown influenced your energy today?            & Text (optional)           &  \\
                                                                              & Work-life balance & How has the lockdown influenced your work life balance today? & Text (optional)           &  \\
                                                                              & Difference        & How different do you feel today was from yesterday?           & 5-Likert scale            &  \\
                                                                              & Difference        & How was today different from yesterday?                       & Text (optional)           &  \\ \bottomrule
\end{tabular}

\label{tab:survey}
\end{table}

\subsection{Collected Data}


The data collection period spanned 21 days. From the 32 recruited participants, we collected 502,485 recordings of smartphone/desktop usage, 1,749 ESM responses, and 265 end-of-day surveys. Table \ref{tab:data statistic} provides an overview of the collected data across devices. To address \textbf{RQ1}, we used both desktop and smartphone data to explore how well-being was during COVID-19 home confinement. For \textbf{RQ2}, due to the limited number of participants contributing desktop data, we focused exclusively on smartphone data and its corresponding ESM responses.

\textit{Smartphone/desktop usage data}. During the data collection, background services of smartphones and desktops continuously tracked the application usage (via \textit{Android Accessibility Services} \footnote{Accessibility Services: \url{https://developer.android.com/reference/android/accessibilityservice/AccessibilityService}}),  notifications (via \textit{Android Notification Listeners} \footnote{Notification Listeners:  \url{https://developer.android.com/reference/android/service/notification/NotificationListenerService}}), as well as the phone's states (only for smartphone, i.e. screen/battery status,  ringer modes, location updates and physical activities). 
Specifically, smartphone app usage was recorded each time an app was opened, while desktop usage was tracked by polling the active application every second. 
For notifications, data captured included arrival time, hashed content (for notification length), sender (private, work, both, or none), and the participant's relationship to the sender (family, friend, work, or none). Table \ref{tab:survey} displays an overview of the collected data for digital device usage, ESM and end-of-day survey. 


\textit{ESM responses}.
There are some popular questionnaires to measure well-being \cite{joseph2004rapid, bradley_measuring_1994}, and the ESM questionnaire adapted from the \textit{Self-Assessment Manikin} (SAM) \cite{bradley_measuring_1994} was employed in this research. In the ESM questionnaire, we collected information on valence, arousal, social role, interruptibility, and tasks from users (see Table \ref{tab:survey}). Specifically, the measurement of items \textit{valence} and \textit{arousal} were adapted from SAM \footnote{Compared to the original SAM questionnaire, we added a time constraint, asking participants to describe their feelings over the past hour.} and rated with a 5-point Likert scale, from \textit{unhappy} to \textit{happy} for valence and from \textit{calm} to \textit{excited} for arousal. These measures were used to represent the users' emotions based on the circumplex model \cite{russell1980circumplex}. The social role, interruptibility, and task questions were sourced from Anderson's assessment framework \cite{anderson2016assessment}.

\begin{figure}[htbp!]
    \centering
    \subfigure[ESM\label{subfig: esm}]{\includegraphics[width=0.265\textwidth]{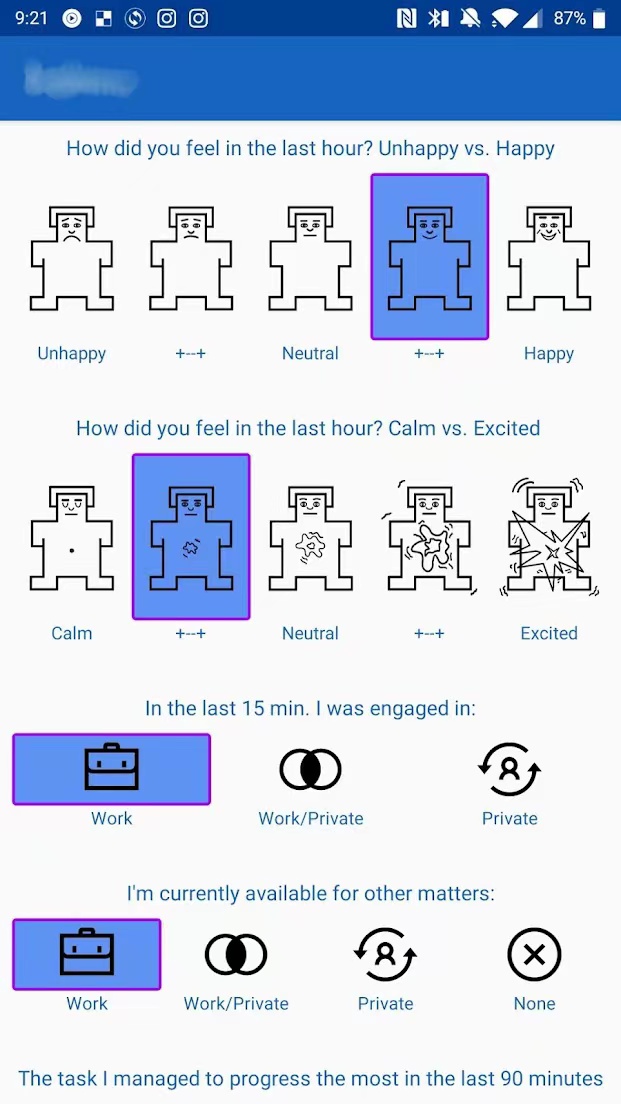}
    }
    \hspace{2cm}
    \subfigure[End-of-day survey \label{subfig: endofday}]{\includegraphics[width=0.25\textwidth]{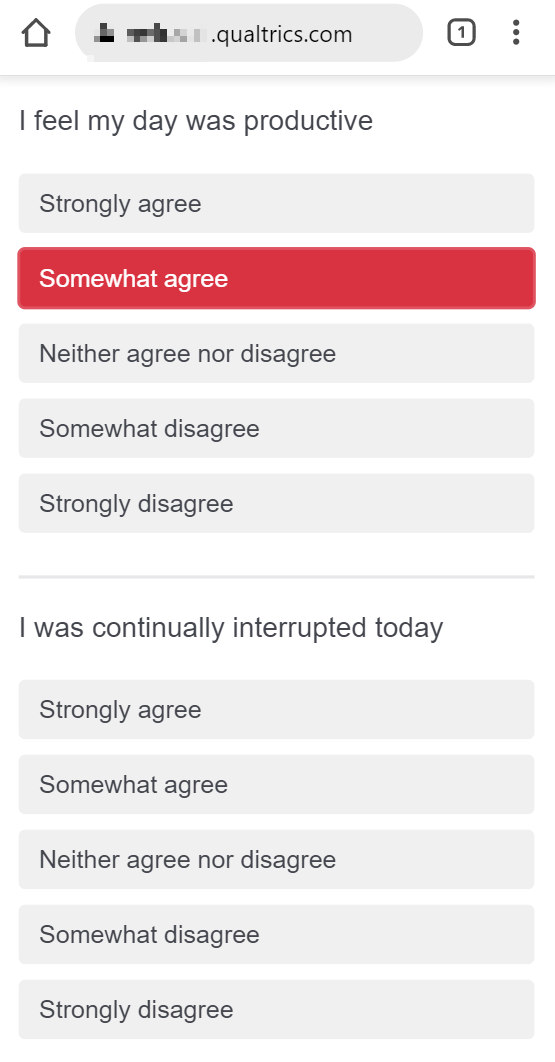}
    }
    \caption{The screenshots for the ESM and End-of-day survey}
\end{figure}

\textit{End-of-day survey}.
As shown in Figure \ref{subfig: endofday}, the end-of-day surveys were optional and included ten items designed to understand participants' well-being during the lockdown, focusing on aspects such as productivity, valence, energy, and work-life balance. Additionally, we surveyed participants on how they felt today compared to yesterday. This information helps describe the dynamics of people's experiences during the COVID-19 lockdown. The end-of-day survey questions were tailored to capture the unique context of our study and were not sourced from existing questionnaires. Each item in the end-of-day survey was rated using a 5-point Likert scale or answered with free text input (see Table ~\ref{tab:survey}). Specifically, items 1 and 2 were rated from 1 to 5, indicating \textit{strongly agree}, \textit{somewhat agree}, \textit{neither agree nor disagree}, \textit{somewhat disagree}, and \textit{strongly disagree}. Items 3, 4, 5, and 9 were rated from 1 to 5, indicating \textit{a great deal}, \textit{a lot}, \textit{a moderate amount}, \textit{a little}, and \textit{not at all}.

\section{Understanding Emotions, Social Roles, and Work-life-balance During Home Confinement (RQ1)}
\label{sec: esm analysis}

In this section, we address \textbf{RQ1} \textit{``How are emotions and social roles during home confinement perceived through self-reports and inferred from app usage patterns?"}. First, we analyse 1522 ESM responses collected via smartphones from 25 participants. Next, we explore the impact of the lockdown by examining the end-of-day survey. Additionally, we explore users' app usage behaviours through unsupervised activity recognition, identifying common activity clusters and their relationship to the perceived psychological metrics.

\subsection{The Perceived Valence, Arousal and Social Roles During Home Confinement}
\label{subsec: rq1-1}

Figure \ref{fig:dis_esm} illustrates the distribution of perceived valence, arousal, and social roles. On average, participants reported a valence of 3.46 (SD = 1.01) and arousal of 2.72 (SD = 1.11), indicating a generally calm and relaxed emotional state. This likely reflects feelings of safety and relief from the external stressors associated with the pandemic, such as the risks of infection.
The absence of daily stressors—like commuting, work pressure, and social obligations—may also contribute to this sense of well-being.
Furthermore, the solidarity and support networks that emerged during the pandemic likely provided additional emotional comfort. However, it is important to note that our data collection occurred during a three-week period as the COVID-19 Stage IV restrictions were lifting.
Had the confinement continued longer, participants likely would have reported more negative emotions, such as boredom or frustration, as they adjusted to the new normal.

\begin{figure}[htbp!]
    \centering

     \subfigure[Valence and arousal\label{subfig: smart}]{\includegraphics[width=0.36\textwidth]{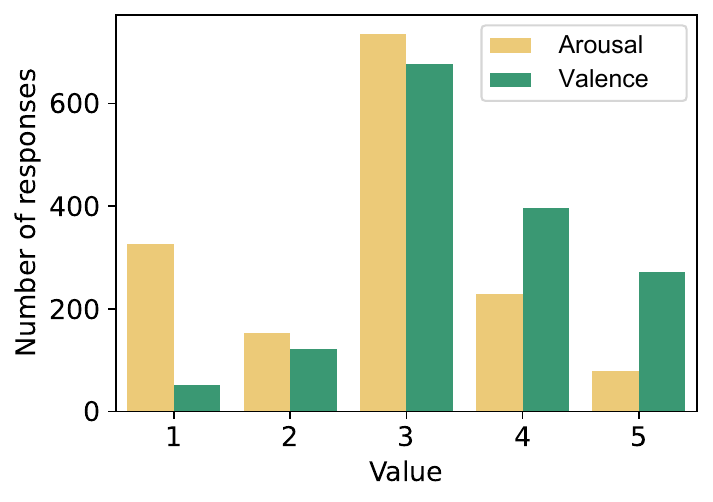}
    }
        \subfigure[Social roles\label{subfig: dis_social_smart}]{\includegraphics[width=0.282\textwidth]{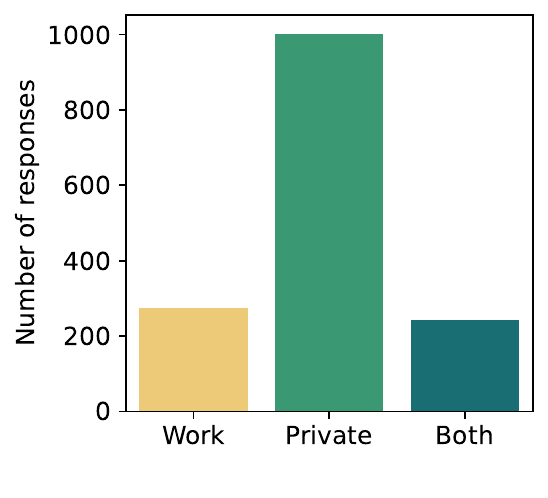}
    }
    \subfigure[Social roles and emotion\label{subfig:social_smart}]{\includegraphics[width=0.27\textwidth]{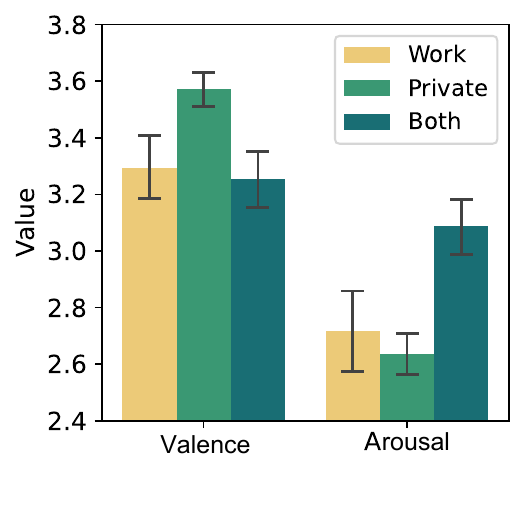}
    }
    \caption{The distribution of perceived valence, arousal, social roles and their relationship}
    \label{fig:dis_esm}
\end{figure}

Next, we investigate the role of social roles in shaping emotional experiences during home confinement. A \textit{social role} refers to the expectations, norms, and behaviours associated with particular social contexts \cite{reise2009item}. These roles usually influence users' behaviours and the type of information they prefer to receive \cite{anderson2016assessment}. As shown in Figure~\ref{subfig: dis_social_smart}, participants predominantly engaged in private-related roles, suggesting individuals tend to prioritize their personal lives and relationships during home confinement.

We further explore how emotions vary across different social roles. Figure \ref{subfig:social_smart} shows that participants reported the highest valence and lowest arousal when engaged in \textit{private}-related roles, compared to \textit{work}-related or hybrid roles. This suggests that personal activities, which are often less stressful and more enjoyable, foster a greater sense of well-being. Conversely, \textit{work}-related and \textit{hybrid} roles were associated with higher levels of stress and pressure. Welch's $t$-tests \cite{zimmerman1993rank} confirm these findings, showing significant differences in valence between \textit{private}-related and the other two roles (i.e., \textit{private} vs. \textit{both}, \textit{private} vs. \textit{work}), with $p<.05$. The findings offer insights into how individuals can structure their daily routines to boost positive emotions during home confinement. Interventions aimed at improving well-being in such situations may benefit from considering the social roles individuals are engaging in \cite{hadjiconstantinou2016web}.

\subsection{The Influence of the Home Confinement on Emotions,  Social Roles and Work-Life Balance}
\label{subsec: rq1-2}

End-of-day survey responses reveal that the lockdown had a noticeable impact on participants' valence, arousal, and work-life balance. As shown in Figure \ref{fig:dis_end}, 54.3\%, 51.3\%, and 54.7\% of responses indicated that the lockdown influenced valence, energy, and work-life balance, respectively. 

In terms of valence, 13 participants reported negative effects, describing feelings of anxiety, sadness, and boredom. Common complaints included missing social interactions (9 responses), the inability to go outside (10 responses), and a sense of boredom (6 responses). For instance,
\textit{``I thought I might be able to see friends, but I did not"}, \textit{``I want to go outside, to the park"}, \textit{``I'm missing my family"}, \textit{``It's hard to say how it's specifically changed my happiness today because it's the build-up of it has gone on for so long. I'm bored and sick of my housemates and it feels like time has stopped"}. 
On the other hand, two participants reported positive effects, such as feeling happier or using the extra time for physical activity and seamless work transitions. 

Regarding arousal, 11 participants indicated a negative impact, expressing feelings of fatigue  (11 responses), lack of motivation (3 responses), or laziness (5 responses), e.g., \textit{``I don't feel motivated to do productive things"}, \textit{``Emotionally tired, perhaps as a result of the lockdown"}. However, two participants noted that they felt more energized during the lockdown.

In terms of work-life balance, the majority of participants reported they faced increased interruptions or distractions (14 responses), they had difficulty separating work from personal life  (11 responses), and they spent more time working due to these disruptions (15 responses), with comments such as \textit{``I worked a bit more than I should've"}, \textit{``Everything blends together, separating work and life is really difficult"}. These findings suggest that the lockdown primarily had a negative impact on participants' work-life balance, with many struggling to manage personal and professional responsibilities. While some respondents reported positive effects, such as spending more time with family or gardening, the overall trend indicates that the lockdown negatively influenced emotional well-being and work-life balance.

\begin{figure}[htbp!]
    \centering
    \subfigure[Productivity\label{subfig:productivity}]{\includegraphics[width=0.28\textwidth]{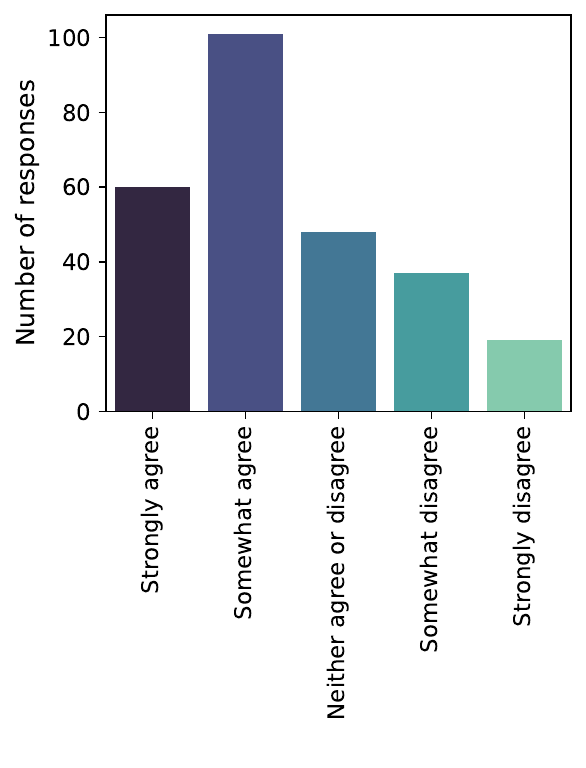}
    }
    \hspace{-0.1cm}
    \subfigure[Interrupted\label{subfig:interrupt}]{\includegraphics[width=0.274\textwidth]{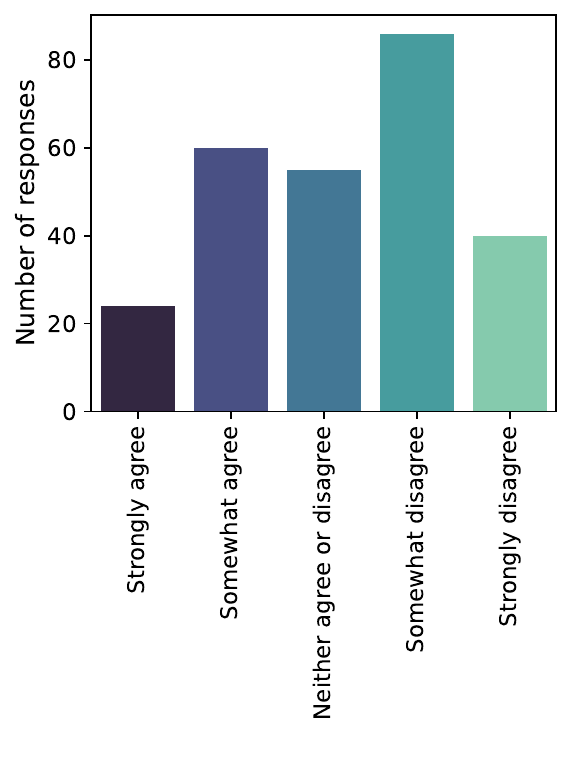}
    }
    \hspace{-0.1cm}
    \subfigure[Influence on well-being\label{subfig:influence}]{\includegraphics[width=0.388\textwidth]{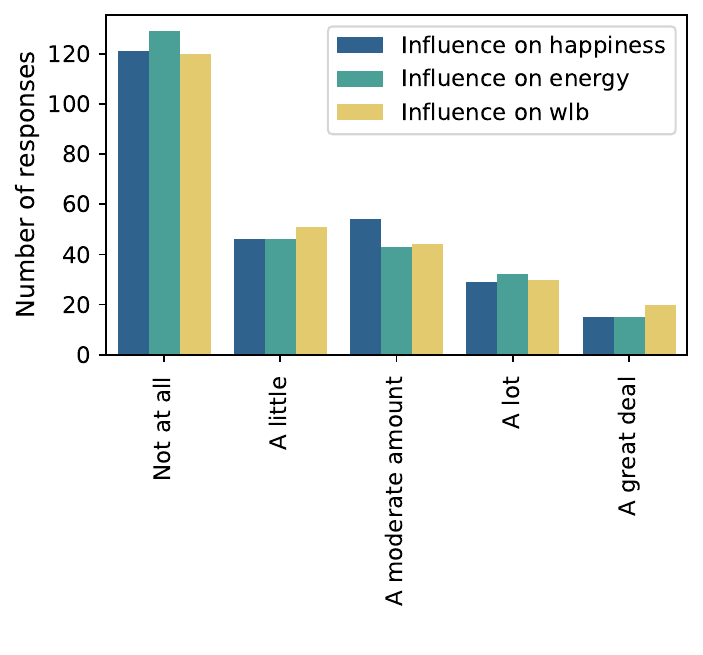}}
    \caption{The distribution of productivity, interruption, and influence during the lockdown}
    \label{fig:dis_end}
\end{figure}

\subsection{Understanding App Usage Behaviours During Home Confinement}
\label{sec: app usage}

\begin{table}[htbp!]
\caption{App categories {(referred to the Google Play Store categories)} and top three frequently used apps. }
\centering
\label{tab:top_apps}
\vspace{5pt}
\scriptsize
\begin{turn}{0}
\begin{tabular}{@{}lllll@{}}
\toprule
\textbf{Upper Category}                  & \textbf{{Subcategory}}        & \textbf{Top App}  & \textbf{Second App} & \textbf{Third App}  \\ \midrule
\textit{Business}                        & Business                 & Teams             & LinkedIn            & Zoom                \\\hline
\textit{Communication}                   & Communication            & Chrome        & Discord             & WeChat              \\\hline
\multirow{12}{*}{\textit{Entertainment}} & Action                   & Among Us      & Among Us            &                     \\
                                         & Casual                   & Scrap Collector   & IdleCourier         &                     \\
                                         & Comics                   & TachiyomiJ2K      &                     &                     \\
                                         & Travel \& Local          & Maps              & PTV                 &                     \\
                                         & Video Players \& Editors & YouTube           & Bilibili            & YouTube Vanced      \\
                                         & Entertainment            & Wow           & Steam           & RainbowSix      \\
                                         & Food \& Drink            & Uber Eats         & Menulog             & Mymacca's           \\
                                         & Health \& Fitness        & Samsung Health    & Nike Training       &                     \\
                                         & Photography              & Photos            & Gallery             & Photos              \\
                                         & Music \& Audio           & Google Play Music & Spotify             & YouTube Music       \\
                                         & Role Playing             & Fate/GO           &                     &                     \\
                                         & Shopping                 & BIGAU             & Gumtree             & Depop               \\\hline

\multirow{2}{*}{\textit{Productivity}}   & Productivity             & Excel         & Outlook         & PowerPoint        \\
                                         & Education                & Canvas Student    &                     &                     \\\hline
\multirow{2}{*}{\textit{Reading}}        & Books \& Reference       & Audible           & AniDroid            & MendeleyDesktop \\
                                         & News \& Magazines        & Joey              & Twitter             & Sync Dev            \\\hline
\textit{Social}                          & Social                   & Facebook          & Instagram           & Snapchat            \\\hline
\textit{Tools}                           & Tools                    & Gboard            & Explorer        & Samsung Keyboard    \\ \hline
\multirow{7}{*}{\textit{Others}}         & Auto \& Vehicles         & Android Auto      &                     &                     \\
                                         & Finance                  & CommBank          & CommSec             & Up                  \\
                                         & House \& Home            & Domain            & Realestate          &                     \\
                                         & Lifestyle                & Samsung Pay       & Tinder              & Hue                 \\
                                         & Medical                  & MyTherapy         & E4 realtime         &                     \\
                                         & Parenting                & FamilyAlbum       &                     &                     \\
                                         & Personalization          & One UI Home       & OnePlus Launcher    & TouchWiz home       \\\bottomrule
\end{tabular}
\end{turn}
\end{table}
 
\subsubsection{App Categorization}

Apps were categorized based on historical usage logs. For \textit{Android} apps, we used the categories from the Google Play Store, excluding uncategorized apps. Desktop applications were manually assigned to equivalent categories from the Google Play Store. Apps with less than one hour of cumulative usage across all participants were excluded.
In total, we identified 28 app categories, which were consolidated into 8 high-level categories.
A summary of the top three most frequently used apps in each category is provided in Table~\ref{tab:top_apps}. Notably, Chrome was categorized under \textit{Communication}, as the Google Play Store has few dedicated browser apps. Categorizing desktop applications posed a challenge, as many are used for work or software development (e.g., \textit{PyCharm}, \textit{WinSCP}).
These were grouped under the \textit{Tools} category, along with utility apps like \textit{Explorer}.

\subsubsection{Unsupervised Activity Recognition From App Sequences}
\label{subsec:unsupervised}
To identify distinct user activities from app usage, we propose an unsupervised approach that combines segmentation and clustering. First, we segment the app usage sequences to capture shifts in activity, then cluster these segments to identify common patterns of behaviour.

\textbf{Segmentation}. To investigate the relationship between app usage and various factors, we propose \textit{App-based Information Gain Temporal Segmentation} (AIGTS), a method that focuses on identifying different activities within app usage. Figure~\ref{fig:igts_cluster_analysis} illustrates how activities are extracted from app usage data. AIGTS is inspired by \textit{Information Gain Temporal Segmentation} (IGTS)~\cite{sadri_information_2017}, an unsupervised method for segmenting multivariate time series. IGTS identifies splits that minimize the weighted entropy across the multivariate time series, where each variable corresponds to whether a specific app category was used.

Unlike IGTS, which focuses solely on the time series, AIGTS detects activity transitions, such as switching from Email or Browser usage (both in the \textit{Communication} category) to browsing Reddit (in the \textit{News and Media} category). This method can segment app usage data at various granularities, ranging from half a day to just 30 seconds. The level of segmentation is determined by the variable $k$, which represents the number of splits made in the time series.
To ensure consistency in segment granularity across days, we compute the optimal $k$ using the following steps: (1) Identify the knee point of the activity data for each day; (2) Calculate the average number of segments per minute for all days based on the knee point; (3) Compute the median segments per minute across all users; (4) Multiply this median by the number of minutes {each user spent on their device} per day and select this value as $k$. 
These steps ensure that the average segment length remains consistent across all days.

\begin{figure}[htbp!]
     \centering
     \includegraphics[width=0.95\textwidth]{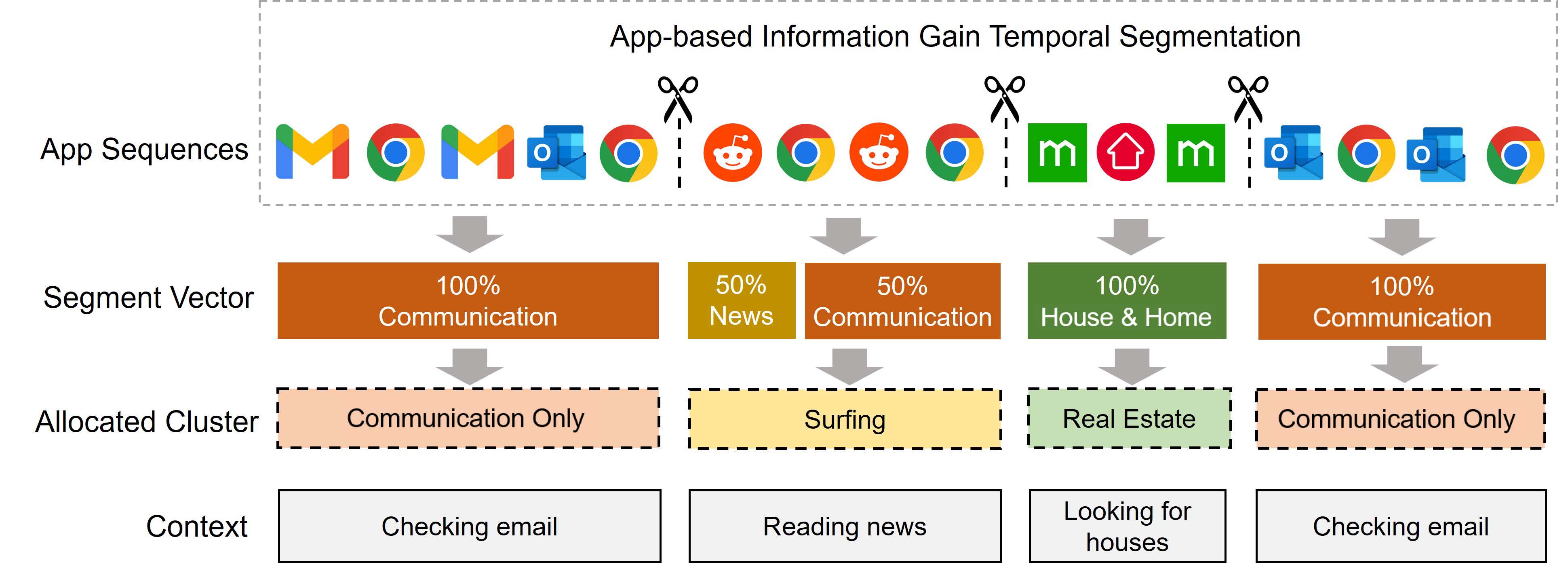}
     \caption{An illustration of unsupervised activity recognition using app usage data. {The \textit{Segment Vector} indicates the app subcategory.}}
     \label{fig:igts_cluster_analysis}
 \end{figure}

\textbf{Clustering}. Next, AIGTS is applied to the all-time series, resulting in a segmented list for each user. To extract features from these segments, \textit{K-means} clustering \cite{likas2003global} was applied. The distribution of different categories appearing in a segment was treated as an $n$ dimensional vector, where $n$ is the number of app categories. To identify the optimal number of segments, \textit{Silhouette Analysis} \cite{lleti2004selecting} was utilized with a maximum of 20 clusters. Especially, the label indicating which cluster a segment belongs to was used as a feature representing the activity within that segment. These segment labels were derived manually from the average distribution of apps in that cluster. In this paper, a segment that belongs to a particular cluster will be referred to as an `activity'.

\begin{figure}[htbp!]
    \centering
    \includegraphics[width=0.68\linewidth]{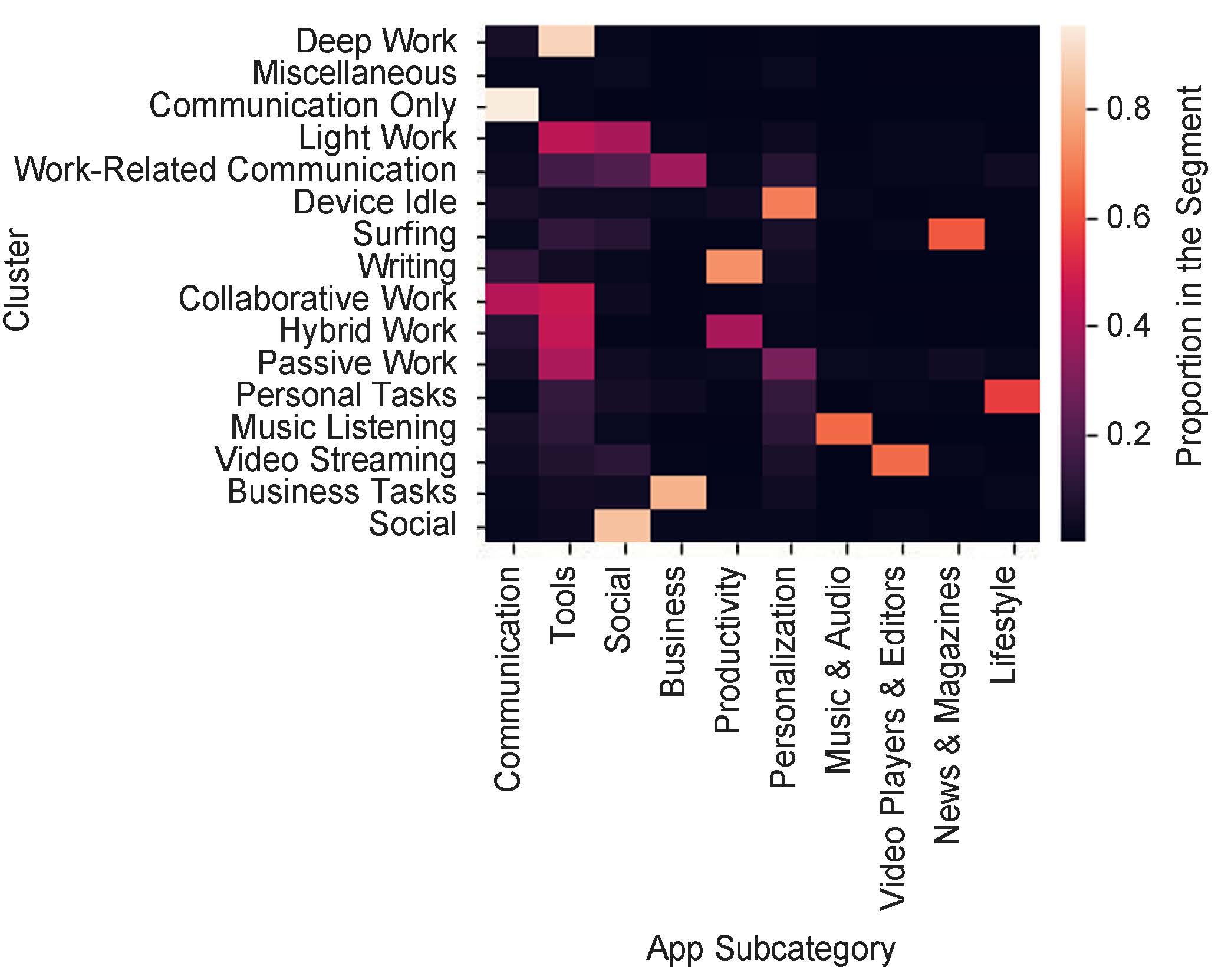}
    \caption{Proportion of app subcategory usage across user activity clusters. Clusters were manually named to describe the distribution of average activities within each cluster.}
    \label{fig:prototype_segments}
\end{figure}

\textbf{Overview of User Activities}.
Following segmentation and clustering, we identified 16 distinct clusters representing user activities, with an average activity duration of 18 minutes.
Figure~\ref{fig:prototype_segments} presents a heatmap showing the proportion of app subcategories in each cluster. The $x$-axis denotes the app categories (categories that infrequently appear in each segment are omitted), while the $y$-axis represents the clusters (manually named such as  \textit{Deep Work}, \textit{Hybrid Work}). The color intensity indicates the relative contribution of each app category to the cluster's overall activity distribution.

The \textit{{Device Idle}} cluster, corresponding to \textit{Personalization} apps (e.g., launchers, as shown in Table~\ref{tab:top_apps}), represents periods when users remain on their device's home screen. We also found that the clusters \textit{Communication Only} and \textit{Device Idle} were the most frequent user activities during the home confinement, accounting for 17\% and 14.6\% of all activities, respectively. These activities are characterized by high proportions of communication-related apps and personalization apps, suggesting users spent considerable time on their devices in passive states or communicating with others. In addition, clusters such as \textit{Work-Related Communication} and \textit{Hybrid Work} exhibited higher proportions of productivity and business apps. These clusters likely reflect professional tasks and remote work activities, which were common during the lockdown period. The differentiation between \textit{Light Work} and \textit{Deep Work} clusters highlights variations in the intensity and type of work-related app usage.

\subsubsection{Correlation Analysis of User Activity/App Usage and Well-Being}.
We further investigate how user activities and app usage behaviors correlate with four key dimensions of well-being: \textit{valence}, \textit{arousal}, \textit{productivity}, and \textit{interruptions}. To assess these correlations, we performed \textit{Spearman Rank Correlation} \cite{de2016comparing} tests between the time spent on different activities/apps and the scores for each dimension. Specifically, \textit{valence} and \textit{arousal} were examined for their relationship to the time spent on user activities in the preceding hour, while productivity and interruptions were calculated based on the end-of-day surveys.

\begin{table}
\centering
\footnotesize
\caption{Top correlations between app usage/user activity and well-being dimensions.}
\label{tab:correlations}
\vspace{5pt}
\begin{tabular}{llll}

\toprule
\textbf{Groups} & \textbf{Variable} & \textbf{Lowest correlation $\rho$} &  \textbf{Highest correlation $\rho$} \\
\midrule
\multirow{4}{*}{\textit{All}} & Valence           &  {Hybrid Work} (-0.07) &              {\textit{Tools} Apps} (0.11)\\
    & Arousal           &            {Business Tasks} (-0.05) &            {\textit{Social} Apps} (0.15)\\
    & Productivity        &        {\textit{Communication} Apps} (-0.24) &  {Work-Related Communication} (0.18) \\
    & Interruptions       &           {Personal Tasks} (-0.15) &            {\textit{Communication} Apps} (0.29)\\\hline
\multirow{4}{*}{\textit{Male}}  & Valence          &  {Communication Only}(-0.08) &  {Work-Related Communication} (0.10) \\
     & Arousal          &              {Device Idle} (-0.12) &      {Communication Only} (0.11) \\
     & Productivity       &       {\textit{Communication} Apps} (-0.15)&  {Work-Related Communication} (0.34) \\
     & Interruptions      &      {Video Streaming} (-0.23) &           {\textit{Communication} Apps} (0.18)\\\hline
\multirow{4}{*}{\textit{Female}}  & Valence        &             Surfing (-0.17) &                   {Miscellaneous} (0.13) \\
    & Arousal        &  {Hybrid Work} (-0.17) &            {\textit{Social} Apps} (0.21)\\
 & Productivity     &        {\textit{Communication} Apps} (-0.30)&               {Personal Tasks} (0.16) \\
 & Interruptions    &           {Personal Tasks} (-0.33) &                  {Device Idle} (0.41) \\\hline
\multirow{4}{*}{\textit{Age 18-24 }} & Valence     &             Surfing (-0.13) &              {\textit{Tools} Apps} (0.09)\\
 & Arousal     &         {\textit{Tools} Apps} (-0.17)&            {\textit{Social} Apps} (0.19) \\
 & Productivity  &        {\textit{Communication} Apps} (-0.21) &      {Hybrid Work} (0.20) \\
 & Interruptions &        {\textit{Social} Apps} (-0.10)&            {\textit{Communication} Apps} (0.16)\\\hline
\multirow{4}{*}{\textit{Age $\geq$25}} & Valence     &  {Communication Only} (-0.12) &              {\textit{Tools} Apps} (0.23)\\
 & Arousal     &  {Hybrid Work} (-0.10) &              {\textit{Tools} Apps} (0.15)\\
 & Productivity  &          {\textit{Tools} Apps} (-0.25)&                  Social (0.20) \\
 & Interruptions &           {Personal Tasks} (-0.24) &            {\textit{Communication} Apps} (0.42)\\
\bottomrule
\end{tabular}
\end{table}

Interestingly, as shown in Table~\ref{tab:correlations}, we found mostly weak correlations, with a few activities exhibiting moderate relationships. For instance, activities related to Tools Apps (e.g., phone home screens and keyboards) showed a moderate positive correlation with valence, especially among older adults. This suggests that certain habitual or passive tasks, such as interacting with productivity tools, may positively influence mood, possibly linked to work-related or routine activities.

Additionally, \textit{Surfing} activity (e.g., browsing news, including pandemic updates) showed a negative correlation with valence among women and the younger age group, indicating that news consumption during this period may be linked to lower mood. The correlation with productivity was notably negative for {\textit{Communication Apps}} (e.g., keyboards, chat apps), but positively correlated with {\textit{Work-Related Communication} activity} (e.g., business apps like Microsoft Teams), highlighting that whether a communication tool is used for professional or personal purposes may influence productivity levels.

Similarly, interruptions were mostly positively correlated with \textit{Communication Apps}, particularly among older adults. This reflects the potentially disruptive nature of non-work-related communication during the day, which may vary by age. Conversely, \textit{Personal Tasks}(e.g., personal apps or leisure activities) were negatively correlated with interruptions, especially for older adults and women, suggesting that non-work-related tasks may help buffer against disruptions, possibly offering a sense of control or relaxation.

\section{Predicting User Emotion and Social Roles from Mobile Usage Data (RQ2)}
\label{sec:results}

In this section, we address \textbf{RQ2} \textit{``Can mobile usage data be used to model emotional well-being during home confinement, and which features are most predictive?"}. First, we introduce the features extracted from mobile usage data and the prediction pipeline. We then list and discuss the prediction results of the general models and individual models. Finally, we examine the significant features for predicting human well-being.

\begin{table}[htbp!]
\caption{The computed features from smartphone usage behaviours. {For the valence model, the features selected by F-statistic $ \geq 5^*$ or $\geq 10^{**}$. For the arousal model, the features selected by F-statistic $ \geq 5^+$ or $\geq 10^{++}$.}}
\label{tab:computed_feature}
\vspace{5pt}
\footnotesize
\begin{turn}{90}
\begin{tabular}{@{}lll@{}}

\toprule

\textbf{Category}                                                                          & \textbf{Feature}        & \textbf{Description}                                                                                     \\\midrule 

\multirow{7}{*}{\textit{\begin{tabular}[c]{@{}l@{}}Contextual\\ Information\end{tabular}}} & weekday$^*$& The day of the week (e.g., Monday, Tuesday)   \\
                                                                                           & time\_period                    & The time of the day (morning, afternoon, evening and midnight) \\
                                                                                           & sin\_weekday$^{**, +}$& Sine Transformation of weekday \\
                                                                                         & cos\_weekday &  Cosine Transformation of weekday \\
                                                                                           & sin\_time$^{**, ++}$&  Sine Transformation of time period \\
                                                                                        & cos\_time$^{**, ++}$&  Cosine Transformation of time period \\
                                                                                           & is\_weekend             & Binary values describing whether is is weekend or not                                                    \\
\hline
\multirow{7}{*}{\textit{Notification}}                                                                    & num\_noti$^{**, ++}$& Number of notifications                                                                                  \\
 & num\_noti\_interacted$^{**, ++}$& Number of notifications interacted \\
 & noti\_cat$^{++}$ & Categorical Transformation of the number of notifications \\
 & noti\_intr\_cat & Categorical Transformation of the number of notifications interacted \\
 & noti\_interacted\_pcg$^{**, ++}$& The percentage of notification interacted to the total number of notifications \\
 & num\_communication & Number of notifications from the Communication Application \\
 & main\_contact$^{**}$ & The main social relationship that contacted with \\
 \hline
\multirow{3}{*}{\begin{tabular}[c]{@{}l@{}}\textit{Physical} \\ \textit{Activity}\end{tabular}}              & nunique\_activity       & Number of unique physical activities                                                                     \\
                                                                                           & main\_activity$^{**, ++}$& The main physical activity of the users                                                                  \\
                                                                                           & num\_activity$^{*, +}$& Number of physical activities                                                                            \\\hline
\multirow{8}{*}{\textit{App}}                                                                             & nunique\_app   & The number of unique apps                                                                                \\
\textit{}                                                                                  & num\_app$^{**, ++}$& The total number of apps                                                                                 \\
                                                                                           & nunique\_app\_cat$^{**}$& The number of unique app categories                                                                      \\
                                                                                           & app\_noti$^{**, ++}$ & The number of notification generated by non-communication apps \\
\textit{}                                                                                  & nunique\_app\_noti$^{**, ++}$& The number of unique apps sent notifications                                                             \\

\textit{}                                                                                  & main\_cat\_app$^{*, ++}$& The main category of apps  \\
                                                    &  second\_cat\_app$^{**, ++}$&                                        The second main category of apps
                                                                             \\\hline
\multirow{5}{*}{\begin{tabular}[c]{@{}l@{}} \textit{Activity Inferred} \\ \textit{from Apps}\end{tabular}      }     & nunique\_user\_activity & The number of unique user activities                                                                     \\
                                                                                           & num\_user\_activity     & The number of user activities                                                                            \\
                                                                                           & main\_user\_activity$^*$& The main user activity                                                                                   \\
                                                                                           & sec\_user\_activity    & The second user activity \\
                                                                                           & entropy\_user\_activity & The entropy of different categories of user activity   
                                                                                           \\ \bottomrule 
\end{tabular}
\end{turn}
\end{table}

\subsection{Feature Extraction} 
Table~\ref{tab:computed_feature} provides an overview of the features utilized in our research. Similar to previous mobile sensing studies \cite{li2022smartphone, does2020zhanna, xu2013preference}, we extract contextual information, notifications-related information, physical activity, app, and activities inferred from apps.
To process the app-usage data, we divide it into segments based on the ESM recording frequency (set at 90 minutes) and store it in temporal sequences. Subsequently, we extract the features from these 90-minute segments. In order to handle cyclical features such as days (e.g., Monday) and times (e.g., Morning), we transform them into sine and cosine values. Additionally, it is worth noting that app usage patterns exhibit variations between weekdays and weekends \cite{katevas2018typical, li2022smartphone}, prompting us to conduct separate analyses for these two categories. Regarding the time period per day, we divide it into six 4-hour segments, i.e., \textit{Early Morning}, \textit{Morning}, \textit{Afternoon}, \textit{Evening}, \textit{Night}, and \textit{Late Night}. Certain frequency-based features, such as notification-related or physical-activity-related features, are converted using quartiles, specifically the minimum, 25th percentile, 50th percentile, and 75th percentile. For instance, the number of interacted notifications is categorized into groups as `0', `1-3', `4-9', and `10+'. Furthermore, the physical activities considered in our study only include active physical activities like bicycling, walking, and running.

To capture user activities related to app usage, we employed our proposed method, AIGTS, as detailed in Section~\ref{subsec:unsupervised}. AIGTS allows us to extract both activities and their corresponding durations. Based on the extracted information, we compute the feature `entropy\_user\_activity' by determining the proportion of each activity's time portion within the entire 90-minute time interval.
In addition to extracting individual features such as `num\_noti', we also generate composite features by combining two existing features. For example, the feature `num\_communication' is calculated by selecting and combining the notification sources. This composite feature counts the number of notifications originating from the \textit{Business}, \textit{Communication}, or \textit{Social} app categories. Notifications that do not belong to these app categories are counted as `app\_noti'. 

\subsection{Prediction Pipeline}
\label{sec:pipeline}

While emotion prediction can be addressed as a classification problem, with emotion levels categorized into two or three classes based on predefined thresholds \cite{pang2002thumbs,moodexplorer}. 
We believe that utilizing regression is a better approach because valence and arousal labels are inherently ordinal data, where the relative order holds significance. In this research, we establish a regression-based pipeline to forecast the emotion scores of participants, as detailed below.

\paragraph{\textbf{Emotion}}
The ESM questionnaires, completed by the participants every 90 minutes, serve as the ground truth in the regression models developed in this study. Specifically, the item `valence' was rated on a 5-point Likert scale from `unhappy' to `happy', where 1 indicates `unhappy' and 5 refers to `happy'. Similarly, another item `arousal' was also rated with a 5-point Likert scale from 1-5, with 1 indicating `calm' and 5 meaning `excited'. To predict the levels of valence and arousal, the app activities within 90 minutes prior to completing the ESM questionnaire were leveraged as features.

\paragraph{\textbf{Regressors}}
In this project, we aim to develop regression models that accurately predict human emotions during home confinement. To accomplish this, we employ three commonly used regression models: \textit{K-Nearest Neighbor} (KNN)~\cite{song2017efficient}, \textit{Random Forest} (RF)~\cite{segal2004machine}, and \textit{Gradient Boosting} (GB)~\cite{friedman2002stochastic}. KNN is a widely utilized model known for its versatility across various domains. It works by finding the K nearest neighbors to a given data point and using their values to predict the target variable. RF and GB, on the other hand, are ensemble models based on decision trees. These models combine predictions from multiple trees. However, there is a distinction in their approach. GB constructs trees or aggregates results in a way that minimizes the loss function, whereas RF constructs individual trees or aggregates results simultaneously.

To explore predictive capacity more comprehensively, we first develop general models using data from all participants combined. Next, recognizing that participants with similar routines or lifestyles may exhibit comparable patterns, we construct group models by dividing participants based on their self-reported habits. Finally, we analyze individual models to account for the unique characteristics and patterns exhibited by different individuals.

\settoheight{\imageheight}{
  \includegraphics[width=0.32\textwidth,keepaspectratio]{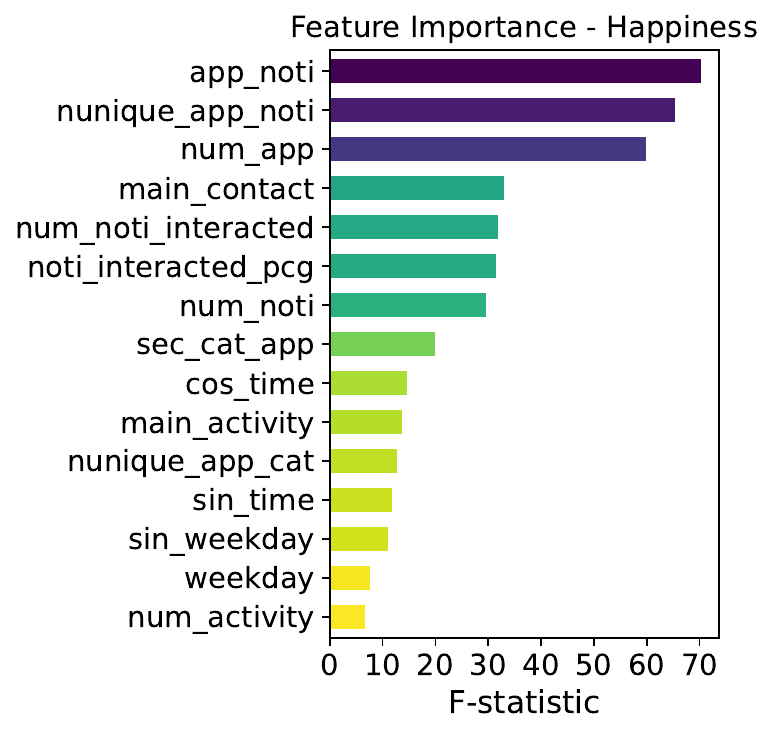}%
}

\paragraph{\textbf{Feature Selection}}
Various methods, such as SHAP~\cite{lundberg2017unified}, LIME \cite{ribeiro2016should}, and the $F$-statistic ($F$) \cite{pope1972use}, can be used for feature selection. In this study, we employ the $F$-statistic to determine the appropriate number of features. While most of our features are numeric, the $F$-statistic measures the univariate linear dependence between each feature and the target variable, taking into account statistical significance. Generally, a higher $F$ value indicates greater importance of the feature in predicting the target variable. In our experiments, we compare regression models using three feature sets: all features, features selected based on an $F$ value greater than or equal to 5, and 
{greater than or equal to 10.}

\paragraph{\textbf{Validation}}

To address the limited richness of our data, we implemented \textit{Nested Cross-Validation} (CV) \cite{muller2016introduction} to train and evaluate our prediction models. Nested CV comprises an outer and an inner loop, both employing \textit{k-fold cross-validation} \cite{wong2019reliable} with different values of $k$. The outer loop splits the data into $k$ folds, using each fold for testing. The inner loop further divides the remaining data into $k$ folds, with each fold serving as a validation set to train the model and optimize hyperparameters. The best parameter settings are selected based on the inner loop results. This approach provides an accurate estimate of the model's unbiased generalization performance.

For the general models, we use a 3-fold CV in the inner loop and a 5-fold CV in the outer loop, repeated over 3 trials. The group models employ 2-fold CV in the inner loop and 3-fold CV in the outer loop. Given the limited data available for individual participants, the individual models use a simpler 5-fold non-nested CV.

\textbf{Baselines {and Metrics}}. Similar to previous human-centered studies \cite{gao2020n, wang2018sensing}, we compare the proposed models with two baselines, \textit{Mean}, which calculates the average score of user responses, and \textit{Median}, which calculates the median value based on the data distribution and treats it as a predicted value. To evaluate the performance of regression models, we employ two commonly used metrics: \textit{Mean Absolute Error} (MAE) and \textit{Root Mean Square Error} (RMSE) \cite{chai2014root}. 
Generally, lower values of MAE or RMSE indicate better prediction performance.

\subsection{Prediction Result}

\subsubsection{General Model} 
After excluding 19 records with invalid ESM data, a total of 1456 app-usage data records were obtained at 90-minute intervals for analysis were used to build the general models.

\begin{table}[htbp!]
\centering
\caption{Prediction performance of the general models for valence and arousal with different feature sets. The regression models are K-Nearest Neighbor (KNN), Random Forest (RF) and Gradient Boosting (GB), compared with Mean and Median baseline. The evaluation metrics are RMSE and MAE (with standard deviation).}
\label{tabl:predict_result}
\vspace{5pt}

\small
\resizebox{\linewidth}{!}{
\begin{tabular}{cccccc}

\toprule

\textbf{RMSE/MAE(SD)} & \textbf{Mean Baseline} & \textbf{Median Baseline} &\textbf{KNN Regressor}                & \textbf{RF Regressor} & \textbf{GB Regressor}     \\

 
 \midrule
 
\multicolumn{6}{c}{\textbf{All features}}    \\ \midrule

\textit{Valence (29)}     &0.986(.002)/0.824(.002) & 1.091(.003)/0.767(.004)& 0.963(.026)/0.776(.022) & \textbf{0.903(.025)/0.717(.021)} & 0.925(.020)/0.743(.012) \\ 

\textit{Arousal (29)}      &1.113(.004)/0.884(.004) & 1.146(.004)/0.784(.004) & 1.088(.023)/0.844(.019)&\textbf{1.005(.028)}/0.766(.020)&1.084(.044)/\textbf{0.756(.037)}\\ 
\midrule

\multicolumn{6}{c}{\textbf{Features with $F$-statistic $\geq5$ }}    \\ \midrule

\textit{Valence (17)}     &0.986(.002)/0.824(.002) &1.091(.003)/0.767(.004) &0.956(.020)/0.769(.018) &\textbf{0.904(.024)/0.717(.021)} &0.986(.020)/0.725(.021)  \\ 

\textit{Arousal (14)}        &1.113(.004)/0.884(.004) & 1.146(.004)/0.784(.004) &1.064(.027)/0.818(.026) &\textbf{1.015(.024)}/0.778(.018) & 1.087(.032)/\textbf{0.747(.022)}  \\  \midrule

\multicolumn{6}{c}{\textbf{Features with $F$-statistic $\geq10$}}    \\ \midrule

\textit{Valence (13)}     &0.986(.002)/0.824(.002) &1.091(.003)/0.767(.004) & 0.936(.020)/0.730(.013) & \textbf{0.912(.024)}/0.725(.018) &0.987(.046)/\textbf{0.724(.019)}  \\ 

\textit{Arousal (12)}       &1.113(.004)/0.884(.004) & 1.146(.004)/0.784(.004) &1.064(.027)/0.819(.026) &\textbf{1.011(.023)}/0.774(.017) &1.076(.035)/\textbf{0.745(.030)}  \\ 
\bottomrule

\end{tabular}
}
\end{table}

The prediction performance for the general models is presented in Table \ref{tabl:predict_result}. The models used in the comparison include Mean baseline, Median baseline, KNN regressor, RF regressor, GB regressor. Among these models, the RF regressor demonstrated superior performance to both baselines in terms of MAE and RMSE for predicting valance and arousal. Interestingly, the RF regressors with different feature sets achieved comparable performance, with all feature sets yielding the best results. This can be attributed to the capability of RF to assign weights to features based on their importance, where incorporating more features leads to improved prediction performance. Furthermore, it is observed that the proposed model generally exhibited better performance in predicting \textit{valence} compared to arousal, the difference in performance may be attributed to the complexity of arousal, which is influenced by various factors such as physiological arousal and cognitive processing. These factors may not be fully captured by mobile usage behaviours alone.

Regarding the \textit{valence} dimension, our results indicate that the RF regressor achieved the best predictive performance across three different feature sets. For the regression model using all features, the RMSE obtained by the RF algorithm was 0.903, surpassing the Median baseline by 17.23\%. Furthermore, the MAE of the RF regressor using all feature sets was 0.717, outperforming the mean baseline by 12.99\% and the median baseline by 6.52\%. Similar results were observed for the models utilizing the top 17 features. Regarding the \textit{arousal} dimension, our results indicate that the RF regressor utilizing all features achieved the best performance in terms of RMSE, with a value of 1.005. This outperformed the Median baseline by 12.30\%. These outcomes suggest that the selected features serve as effective predictors of \textit{arousal} for participants during home confinement.

\subsubsection{{Group Models}}
\label{subsec:prediction_group}

We gathered participant work start times from pre-study questionnaires and categorized participants into two distinct groups: \textit{EARLY} Workers (678 data instances from 12 participants who typically commence work between 7:00 and 10:00 AM) and \textit{LATE} Workers (778 data instances from 13 participants who begin work after 3:00 PM). For the early workers, the average self-reported \textit{valence} is 3.2 ($\pm0.9$) and \textit{arousal} is 2.9 ($\pm1.0$). For the late workers, the average \textit{valence} is 3.7 ($\pm1.0$) and \textit{arousal} is 2.6 ($\pm1.2$). Prediction models were subsequently developed for each group using all features.

As shown in Figure~\ref{subfig:valence_result_rmse_group}, when predicting \textit{valence} for early workers, the models performed similarly to the baselines. However, for late workers, the RF regressor outperformed the Mean baseline, achieving a 13.82\% reduction in RMSE and a 20.36\% reduction in MAE. A similar pattern is observed in Figure~\ref{subfig:arousal_result_rmse_group}. For \textit{arousal} prediction, the models demonstrated limited predictive ability for early workers but performed better for late workers, with the RF regressor reducing RMSE by 17.88\% and MAE by 27.74\% compared to the Mean baseline. Compared to the general models, there was no improvement in \textit{valence} prediction for early workers, whereas for late workers, the models showed slight improvements (5\% in RMSE) in both \textit{valence} and \textit{arousal} predictions.

\begin{figure}[htbp!]
    \centering
    \subfigure[RMSE for valence\label{subfig:valence_result_rmse_group}]{\includegraphics[height=0.84\imageheight]{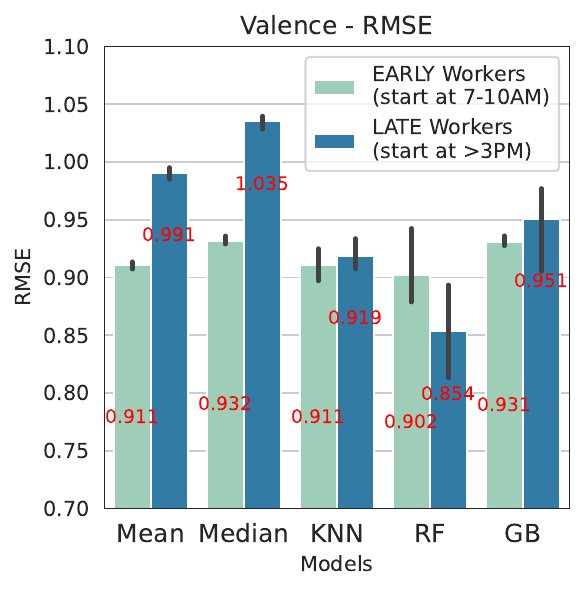}}
    \hspace{-.7em}
    \subfigure[MAE for valence\label{subfig:valence_result_group}]{\includegraphics[height=0.84\imageheight]{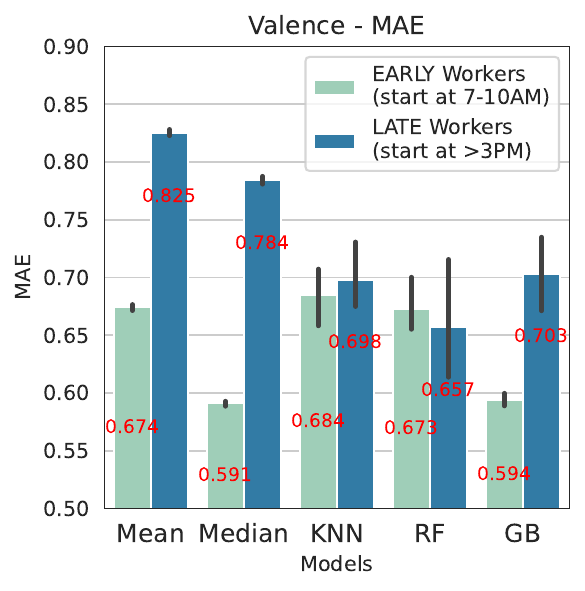}}
    \hspace{-.7em}
    \subfigure[RMSE for arousal\label{subfig:arousal_result_rmse_group}]{\includegraphics[height=0.84\imageheight]{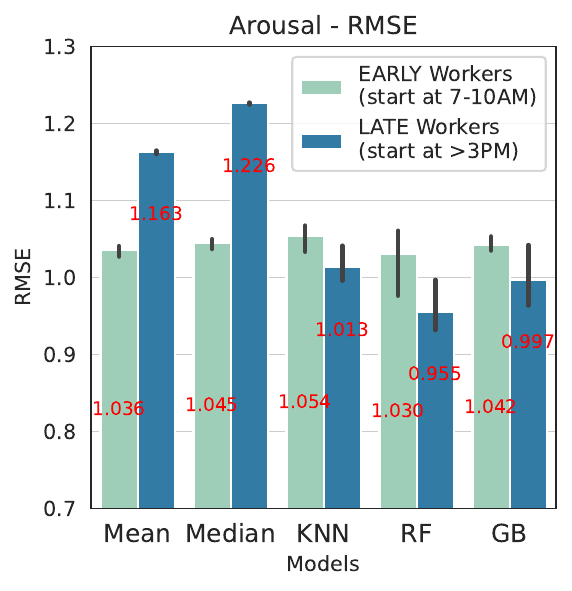}}
    \hspace{-.7em}
    \subfigure[MAE for arousal\label{subfig:arousal_result_group}]{\includegraphics[height=0.84\imageheight]{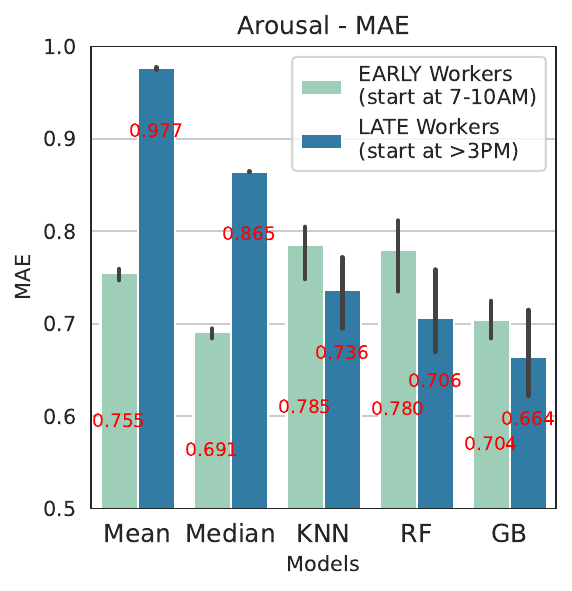}}
   \caption{Prediction performance of \textit{valence} and \textit{arousal} models for two user groups: \textit{EARLY} workers (typically starting work between 7:00 and 10:00 AM) and \textit{LATE} workers (starting work after 3:00 PM). The models are evaluated with 3-fold cross-validation, where the error bar presents the standard deviation.}
    \label{fig:group_prediction}
\end{figure}

These results highlight the potential value of considering user groups and routines when studying emotions. Specifically, the findings suggest that early workers' emotions may be influenced by factors beyond app usage, such as external environmental cues during the day. In contrast, late workers may exhibit a stronger link between app usage behaviours and emotions. Further insights will be explored in the later section (Section~\ref{sec:feature_importance}) by analyzing feature importance.

\subsubsection{Individual Models}

The individual models for both \textit{valence} and \textit{arousal} were trained using all 29 features. Due to insufficient data for some participants, only those with over 50 data points were included, resulting in a total of 13 participants and 1,220 data points.
The prediction results for these models, with the ID of participants indicated on the y-axis, can be seen in Figure~\ref{fig:individual_prediction}. Overall, we can observe variations in the prediction performance of \textit{valence} and \textit{arousal} across different participants. For instance, participant `M11OAR' consistently exhibits higher MAE and RMSE on both \textit{valence} and \textit{arousal} than the other participants, e.g., `H12DPB' and  `E02OKG'. This variability suggests that individual differences play a significant role in determining the accuracy of the predictions, highlighting the need to consider personalized factors when developing prediction models for emotion in home confinement situations.

Additionally, it is worth noting that while the proposed \textit{valence} models exhibited better prediction performance than baseline models for certain participants (e.g., `M02MMP', `L10FGB'), there were participants who did not achieve satisfactory results. The possible reasons may be two-fold: firstly, individual differences play a crucial role in determining \textit{valence}, and it may not be straightforward to accurately predict \textit{valence} solely based on mobile usage behaviours for some participants. This highlights the complexity and subjectivity of \textit{valence} and the need for a more nuanced understanding of individual emotional experiences; secondly, insufficient data availability may have limited the performance of the predictive models for some participants. Having a larger and more diverse dataset would enable better training and improve the performance of the predictive models. These findings contribute to our understanding of personalized prediction models for well-being, which could ultimately lead to more tailored interventions and support systems to enhance individuals' well-being during periods of home confinement.

\begin{figure}[htbp!]
    \centering
    \subfigure[valence (MAE)]{\includegraphics[height=0.9\imageheight, trim={ 0 0 0 19px}, clip]{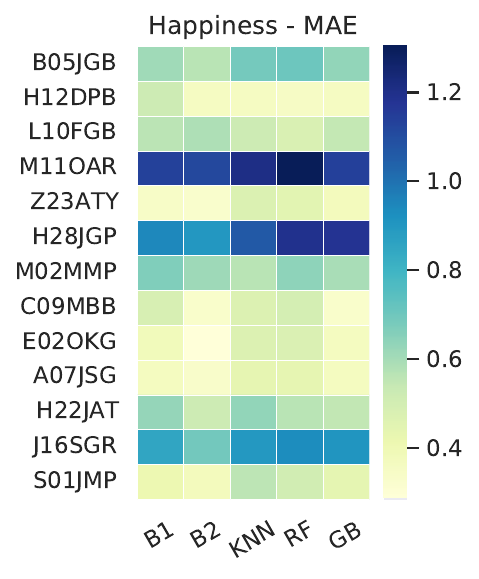}}
    \subfigure[valence (RMSE)]{\includegraphics[height=0.9\imageheight, trim={ 0 0 0 20px}, clip]{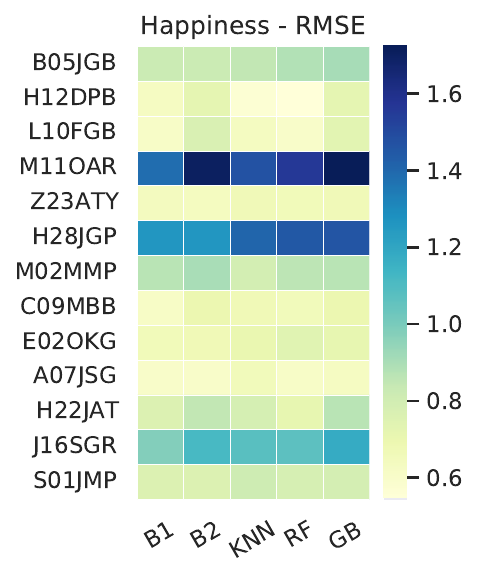}}
    \subfigure[arousal (MAE)]{\includegraphics[height=0.9\imageheight, trim={ 0 0 0 19px}, clip]{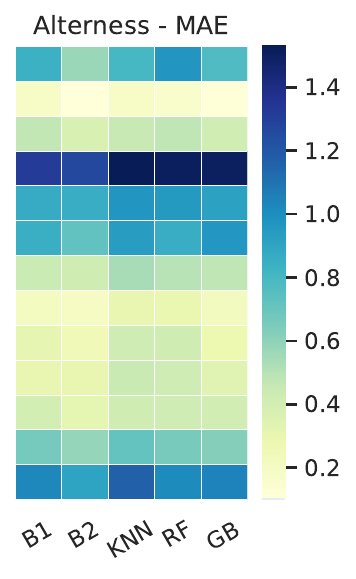}}
    \hspace{2em}
    \subfigure[arousal (RMSE)]{\includegraphics[height=0.9\imageheight, trim={ 0 0 0 20px}, clip]{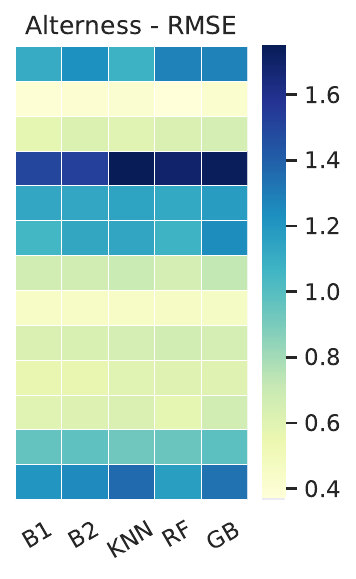} }
    \caption{Prediction performance of individual models using all features}
    \label{fig:individual_prediction}
\end{figure}

\subsection{Feature Importance} 
\label{sec:feature_importance}

To interpret the key factors influencing the prediction of \textit{valence} and \textit{arousal} during home confinement, we examine the feature importance of the best-performing models, which in this case are the RF regressors. The RF regressors utilize impurity-based feature importance, which measures how much including a particular feature reduces impurity across the decision tree nodes. A higher score indicates a more important feature, as it leads to a greater reduction in impurity. For brevity, we focus on the top 15 features for the general model and compare them with the top 5 features for the group models.

\begin{figure}[htbp!]
    \centering
    \begin{subfigure}
       {\includegraphics[height=1.55\imageheight]{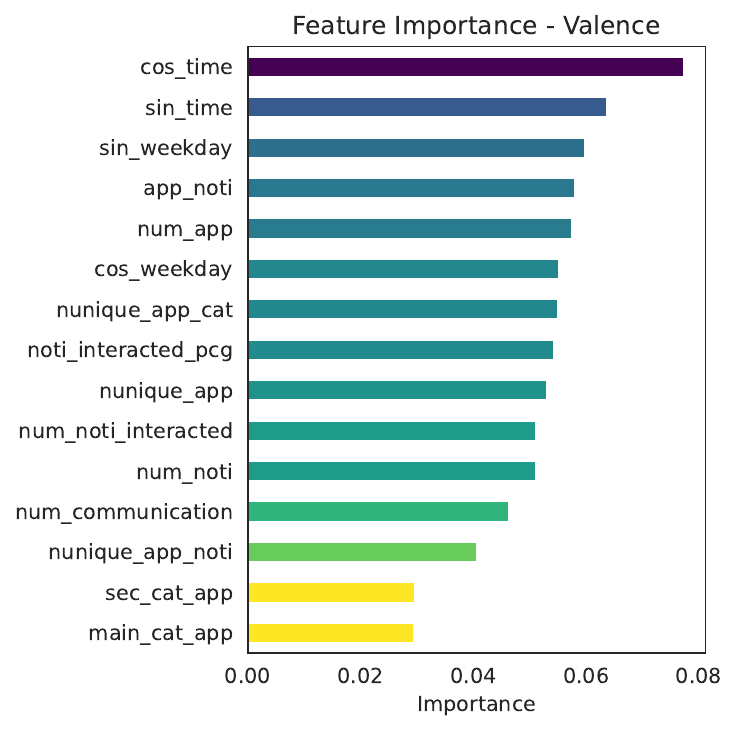}}
    \end{subfigure}
    \begin{subfigure}
       {\includegraphics[height=1.55\imageheight]{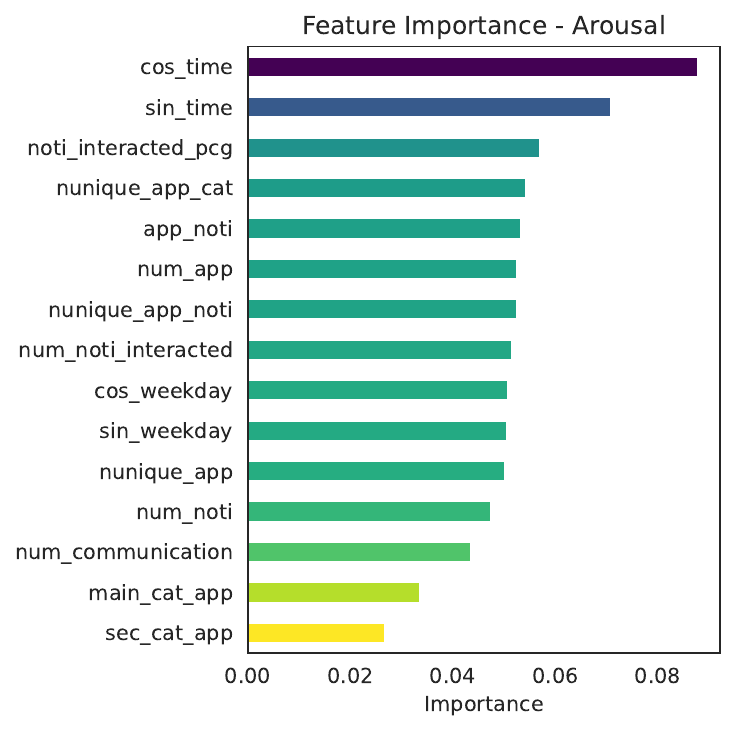}}
    \end{subfigure}
    \caption{Top 15 features based on impurity score with RF regressors of the general models.}
    \label{fig:feature_importance}
\end{figure}

As depicted in Figure~\ref{fig:feature_importance}, we observe that the most important features for both \textit{valence} and \textit{arousal} are similar, but with different weights. These features primarily include time-related, app-category, and notification features. 

Both general models consistently identify `sin\_time' and `cos\_time' as the top two important features, indicating that the time period of the day greatly influences participants' \textit{valence} and \textit{arousal}. 
Additionally, the day of the week (`weekday') impacts both models, although it has less importance in the \textit{arousal} model. 

Surprisingly, the feature `is\_weekend' does not appear among the top 15 features in the general models, but it appears as important in each group model. It is worth mentioning that `time\_period' is the most important feature for predicting \textit{valence} for \textit{EARLY} workers, but not for \textit{LATE} workers.
These findings differ slightly from the observations reported in \cite{katevas2018typical}, which found that app usage patterns varied greatly by time and day. The discrepancy may be attributed to the unique circumstances of home confinement, where individuals may experience a loss of time perception and a blurring of the distinction between weekdays and weekends. This aligns with the findings presented in \cite{cellini2020changes}, which suggest that the perception of time and daily routines can be disrupted during periods of confinement.

\begin{figure}[htbp!]
    \centering
    \begin{subfigure}
        {\includegraphics[width=0.49\textwidth]{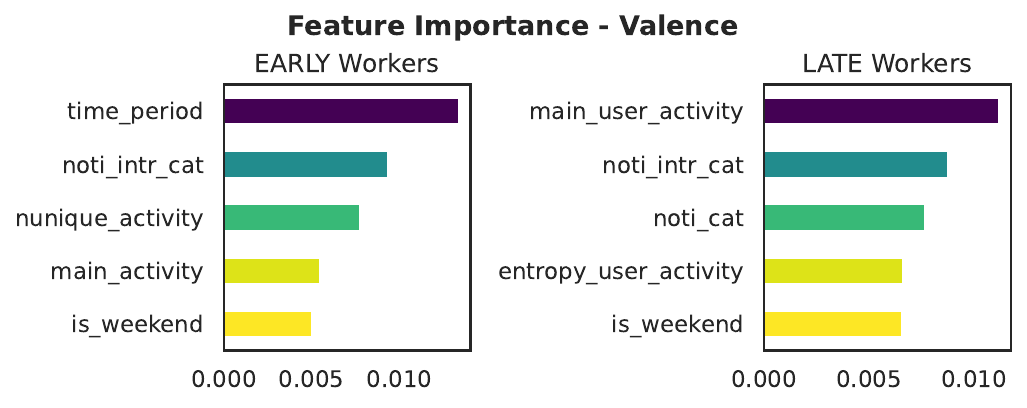}}
    \end{subfigure}
    \hfill
    \begin{subfigure}
        {\includegraphics[width=0.49\textwidth]{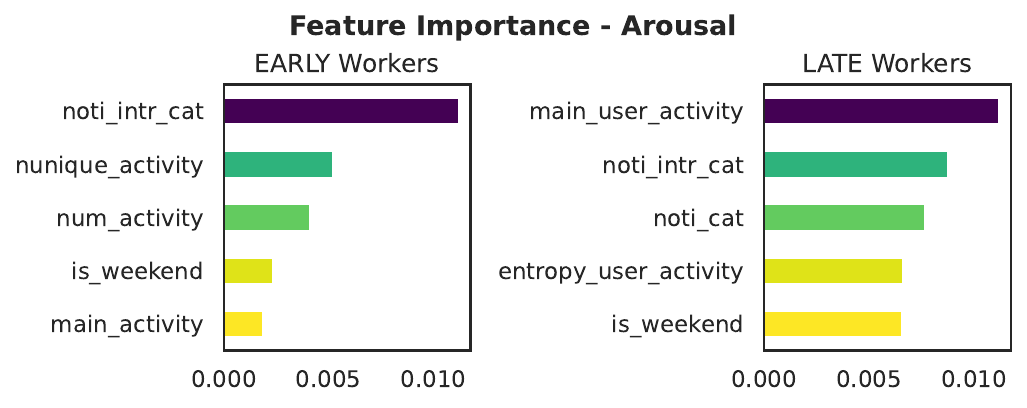}}
    \end{subfigure}
    \caption{Top 5 features based on impurity score with RF regressors of group models.}
    \label{fig:feature_importance_group}
\end{figure}

Besides, as shown in Figure~\ref{fig:feature_importance_group}, for \textit{LATE} workers, the same five features with similar importance distributions are observed for both \textit{valence} and \textit{arousal} predictions, all are related to app interaction. In contrast, \textit{EARLY} workers show a more diverse feature set; while the interacted app notification is the second most important, physical activity also contributes to \textit{valence} prediction. This difference may reflect the broader range of daytime activities available to \textit{EARLY} workers, potentially explaining their lower prediction performance discussed in Section~\ref{subsec:prediction_group}. For \textit{LATE} workers, the reliance on app interaction highlights a potential concern regarding phone dependency.

\section{Discussion and Implications}\label{sec:dis}

This study explores the impact of home confinement on individuals' well-being, social roles, and work-life balance, using unobtrusive mobile phone sensing to gain insights into participants' emotional states and behaviors. Although the study specifically focuses on the COVID-19 lockdown, we argue that its findings have broader relevance, especially in light of the growing trend of remote work and other situations of enforced isolation.

\subsection{Well-being and Social Roles During Home Confinement (RQ1)}

Our analysis revealed that participants generally experienced a state of calmness and relaxation, reflected in relatively high levels of valence (positive emotional state) and lower levels of arousal (indicating a lack of intense emotional stimulation). While this suggests a degree of emotional stability during the lockdown, the study also highlights the negative effects of home confinement on work-life balance. Participants reported longer work hours and the blurring of personal and professional boundaries, which resonates with the challenges faced by remote workers in post-pandemic contexts. The segmentation of app usage data also identified key behaviors, such as increased use of communication apps and idle device time, which may reflect both increased virtual socialization and available free time.

In wider contexts, home confinement is not limited to pandemic scenarios but is increasingly relevant to other forms of isolation or reduced mobility, such as remote work, post-operative recovery, disability-related confinement, or even digital addiction. The growing trend of telework and flexible work arrangements in the post-pandemic world further underscores the importance of understanding the emotional and behavioral impacts of prolonged confinement in various forms.

\subsection{Predicting Emotional States Through Mobile Usage (RQ2)}
In response to RQ2, we developed regression models that predict participants' valence and arousal based on their mobile usage data. These models demonstrated strong predictive performance, emphasizing the value of mobile phones as a tool for tracking emotional well-being in real time. Our findings indicate that time-related features (e.g., the timing of mobile phone usage) were key predictors of both valence and arousal, followed by app notifications and interactions with those notifications. This temporal sensitivity suggests that personalized interventions based on users' specific routines could enhance emotional monitoring and support during home confinement.

One key finding was the variability in the performance of prediction models across different user groups. For example, users who start work later in the day relied more heavily on mobile phone interactions, while those with early start times would benefit from models that incorporate data from other activities, such as physical exercise or interactions with environmental cues. This insight is crucial in extending the applicability of our findings to various forms of isolation beyond the pandemic. Whether it's remote workers, individuals recovering from surgery, or people experiencing limited mobility due to health conditions, the need for individualized and context-aware models remains a central takeaway.

\subsection{Generalizability of the Findings}
Although this study specifically focuses on the COVID-19 home confinement context, we believe the findings can extend to a range of other isolation scenarios. While the emotional triggers -- such as fear of infection in a pandemic or frustration with limited mobility in recovery -- may differ, the \textbf{core psychological mechanisms} and \textbf{emotional dynamics} of isolation are broadly similar. Our study observed patterns like emotional detachment, stress, and blurred work-life boundaries, all of which are commonly experienced during prolonged isolation, regardless of the specific circumstances.

The emotional dynamics of isolation -- such as changes in social interactions, work-related stress, and adjusting to new routines -- are consistent across scenarios like remote work, post-surgical recovery, or other forms of confinement. For example, both remote workers and those recovering from surgery face challenges with work-life balance and social isolation. Our findings about app usage—especially productivity apps promoting positive emotional states—are relevant in these contexts, suggesting that digital tools can support well-being across various isolation scenarios.

Additionally, the mobile sensing approach used in this study, which tracks behaviors like app usage and interaction timing, can be applied beyond the studied context. While the types of apps and behaviors may vary, the principle of using mobile devices to track emotional states in real-time is adaptable to different situations, such as remote work or recovery periods. The ability to monitor and intervene based on individual usage patterns could be valuable in a wide range of confinement contexts.

Lastly, context-aware interventions are crucial for emotional well-being during isolation. Our study highlights the importance of personalized, real-time interventions—whether it's through managing digital interruptions, promoting relaxation, or encouraging physical activity. These strategies can be adapted for any form of isolation, from remote work to recovery, offering broader applicability for improving mental health.

\section{Limitations}\label{sec:lim}
This study has several limitations that must be acknowledged, which may affect the validity of our findings \cite{trochim2001research}:

\textbf{Time and Geographical Constraints (Threats to Construct and External Validity)}.
The timing and location of data collection limit the generalizability of our results. We could not compare users' well-being and digital device usage before and during the COVID-19 lockdown due to the unpredictability of lockdown policies. Although participants reported their experiences via self-reported questionnaires, the effects of different lockdown measures and their duration may vary across regions. As the study was limited to Melbourne, the findings may not fully reflect experiences in other areas with differing restrictions.

\textbf{Short Duration of Data Collection (Threats to Internal Validity).}
The data collection period was limited to three weeks to minimize participant burden during home confinement. While this yielded a substantial dataset, a longer observation period would have provided deeper insights into how digital device usage and emotional well-being evolve over time, particularly as individuals adapt to prolonged lockdowns. Future studies should consider longer data collection windows to capture these potential changes.

\textbf{Device Usage and Untracked Data (Threats to Internal and Construct Validity).}
A key limitation is the potential for incomplete data due to participants using untracked devices, such as tablets or additional smartphones, or even desktops, while only contributing smartphone data. These unmonitored interactions could have influenced emotional states and introduced noise into the data. The absence of comprehensive tracking across all devices limits our ability to fully assess participants' total digital behavior and emotional responses, suggesting the need for more extensive device tracking in future studies.

\textbf{Sample Representativeness (Threats to External and Construct Validity)}.
The sample size (32 participants) and recruitment methods (via Facebook and Discord ads) may have resulted in a skewed population, likely favoring tech-savvy individuals, particularly information workers. Additionally, participants who felt more positively about being at home during the lockdown may have been overrepresented. These factors limit the generalizability of the findings to a broader population. However, despite these challenges, the data collected was sufficient for analysis.

\endgroup

\section{Conclusion}\label{sec:conclusion}

In this field study, we explored how human emotions, social roles, and mobile usage behaviors were affected during the home confinement period. This rare event offers optimal conditions while minimizing confounding factors like nature and physical social interactions. It contributes to understanding the mental health implications of heavy reliance on technology and work-from-home arrangements. By tracking mental well-being, the study highlights how individuals manage work-life balance and adapt their schedules or arrangements (e.g., work-from-home or half-office half-home) effectively. Additionally, it provides valuable insights into psychological states and smartphone usage behaviors during confinement, informing interventions for similar scenarios, such as post-operative home rehabilitation. Understanding how people navigate home confinement and the role of technology in their lives can guide the design of smart environments and remote support programs, improving care quality. These findings also have broader implications for understanding human behavior during crises, supporting the development of policies and interventions for future challenges.

Despite the challenges we faced in collecting data during the lockdowns, we believe that gaining insights into human well-being and mobile usage behaviors during home confinement is crucial. This understanding is not only valuable for policymakers in implementing effective measures during future outbreaks, but also for individuals to become more self-aware of their well-being and mental health. Moving forward, our research will expand to investigate users' desktop usage behavior. By comparing the differences in usage behavior across different platforms, we aim to gain a deeper understanding of people’s daily lives and work-life balance.

\section{Acknowledgments}
We sincerely thank the anonymous reviewers. We state that this manuscript has no relation to our prior publications (or concurrently submitted papers).

\newpage
\bibliographystyle{elsarticle-num-names}
\bibliography{00-bibliography}

\end{document}